\documentclass{article}
\usepackage{amssymb,amsmath}
\usepackage{epsfig}
\usepackage{pdflscape}
\usepackage{subfigure}
\usepackage{color}
\usepackage{lscape}
\usepackage{bbm}
\usepackage{bm}
\usepackage{mathrsfs}
\usepackage{amsfonts}
\usepackage{graphics}
\usepackage{indentfirst}
\usepackage[square,numbers,sort&compress]{natbib}
\usepackage{multicol}

\newtheorem{rem}{Remark}[section]
\newcommand{\br}{\begin{rem}}
\newcommand{\er}{\end{rem}}
\newtheorem{ex}[rem]{Example}
\newcommand{\bex}{\begin{ex}}
\newcommand{\eex}{\end{ex}}
\newtheorem{Def}[rem]{Definition}
\newcommand{\bd}{\begin{Def}}
\newcommand{\ed}{\end{Def}}
\newtheorem{theorem}[rem]{Theorem}
\newcommand{\bt}{\begin{theorem}}
\newcommand{\et}{\end{theorem}}
\newtheorem{Prop}[rem]{Proposition}
\newcommand{\bp}{\begin{Prop}}
\newcommand{\ep}{\end{Prop}}
\newtheorem{lemma}[rem]{Lemma}
\newcommand{\bl}{\begin{lemma}}
\newcommand{\el}{\end{lemma}}
\newcommand{\be}{\begin{equation}}
\newcommand{\ee}{\end{equation}}
\newcommand{\bea}{\begin{eqnarray}}
\newcommand{\eea}{\end{eqnarray}}
\newcommand{\pa}{\partial}
\newcommand{\nn}{\nonumber}
\newcommand{\adots}{\mathinner{\mkern2mu\raise1pt\hbox{.}\mkern2mu
\raise4pt\hbox{.}\mkern2mu\raise7pt\hbox{.}\mkern1mu}}

\title{Stationary Coupled KdV Hierarchies \\  and Related Poisson Structures}

\author{Allan P. Fordy\thanks{School of Mathematics,
University of Leeds, Leeds LS2 9JT, UK. ~~E-mail: a.p.fordy@leeds.ac.uk}
$\,$ and Qing Huang\thanks{School of Mathematics, Center for Nonlinear Studies, Northwest University, Xi’an 710069,
People’s Republic of China ~~E-mail: hqing@nwu.edu.cn}
}

\begin{document}

\maketitle

\begin{abstract}
In this paper we continue our analysis of the stationary flows of $M$ component, coupled KdV (cKdV) hierarchies and their modifications.  We describe the general structure of the $t_1$ and $t_2$ flows, using the case $M=3$ as our main example.  One of our stationary reductions gives $N$ degrees of freedom, superintegrable systems.   When $N=1$ (for $t_1$) and $N=2$ (for $t_2$), we have Poisson maps, which give multi-Hamiltonian representations of the flows.  We discuss the general structure of these Poisson tensors and give explicit forms for the case $M=3$.  In this case there are 3 modified hierarchies, each with 4 Poisson brackets.

The stationary $t_2$ flow (for $N=2$) is separable in parabolic coordinates.  Each Poisson bracket has rank 4, with $M+1$ Casimirs.  The $4\times 4$ ``core'' of the Poisson tensors are nonsingular and related by a ``recursion operator''.  The remaining part of each tensor is built out of the two commuting Hamiltonian vector fields, depending upon the specific Casimirs.  The Poisson brackets are generalised to include the entire class of potential, separable in parabolic coordinates.  The Jacobi identity imposes specific dependence on some parameters, representing the Casimirs of the extended canonical bracket.  This general case is no longer a stationary cKdV flow, with Lax representation.  We give a recursive procedure for constructing the Lax representation of the stationary flow for all values of $M$, {\em without} having to go through the stationary reduction.
\end{abstract}



\section{Introduction}

In this paper our main interest is the construction of integrable and superintegrable finite dimensional Hamiltonian systems (see \cite{90-16,90-22,13-2,14-2}) with multi-Hamiltonian representations and possessing a Lax pair.  One such construction is to consider the stationary reductions of completely integrable nonlinear evolution equations (PDEs of soliton type).
The observation that the stationary flows of the KdV hierarchy are themselves integrable, finite dimensional Hamiltonian systems was made by Bogoyavlenskii and Novikov in 1976 \cite{76-5}.  These were later shown to be bi-Hamiltonian in \cite{f87-3} (see also \cite{00-4}).  Stationary flows of other ``soliton hierarchies'' were discovered in \cite{f91-1,f95-3}.

In \cite{f23-1} we revisited this subject and found bi- and tri-Hamiltonian representations of a number of superintegrable systems.  These were associated with the first two nontrivial flows of the $M$ component cKdV hierarchies, related to the {\em energy-dependent Schr\"odinger equation} (see  \cite{f87-5,f89-2}).  In \cite{f23-1} we concentrated on the cases $M=1$ and $M=2$, corresponding to the KdV and Dispersive Water Wave (DWW) hierarchies respectively.
In one particular coordinate system (associated with a squared eigenfunction substitution) these flows share interesting common features, which we emphasise in this paper.   These features could already be seen in the KdV and DWW hierarchies, but are further illustrated here for the $M=3$ case.

In particular, the stationary $t_1$ flow has a Hamiltonian function of the form
\begin{subequations}
\be\label{gen-t1flow}
h^{(Q)} = \frac{1}{2} \sum_{i=1}^N \left(P_i^2+\frac{2 \beta_i}{Q_i^2}\right) + U\left({\bf Q}^2\right),
\ee
where ${\bf Q}^2 = \sum_1^N Q_i^2$, and is superintegrable, with a {\em universal} collection of first integrals.  These systems are known to be superintegrable (see \cite{90-22} for the case $N=3$ and \cite{09-10} in general), but for the particular choices of $U\left(\sqrt{{\bf Q}^2}\right)$ which appear in this paper (which are specific polynomials of ${\bf Q}^2$), they have a Lax formulation, as follows from being stationary reductions.  For the case $N=1$, we also give a multi-Hamiltonian formulation.

The stationary $t_2$ flow has a Hamiltonian function of the form
\be\label{t2-conc}
h^{(Q)} = \frac{1}{2}P_1^2+ \frac{1}{2} \sum_{i=2}^N \left(P_i^2+\frac{2\beta_i}{Q_i^2}\right) + U\left(Q_1,\sum_{i=2}^N Q_i^2\right),
\ee
\end{subequations}
which has the same universal set of integrals as (\ref{gen-t1flow}), but for $2\leq i<j\leq N$, as well as an integral $f^{(Q)}$, which survives the reduction to the case $N=2$.  This case generalises the well known Hamiltonians, which are separable in parabolic coordinates (see Equation 2.2.41 in \cite{90-16}).  The particular potentials which arise as stationary flows belong to the polynomial sequence and, as a consequence, these have a Lax formulation.  Again, for the case $N=2$, we also give a multi-Hamiltonian formulation.  Furthermore, this multi-Hamiltonian formulation can be extended to the entire class of potentials, separable in parabolic coordinates.

In Section \ref{sec:Lax-background} we give a brief overview of cKdV hierarchies, giving some additional details for the $M=3$ case.  We also give some background on Miura maps and some general constructions of stationary flows.  In particular, we derive two different representations of stationary flows, using the {\em first} and {\em second} Hamiltonian structures to respectively define coordinates $\bf q$ and $\bf Q$, which play an important role in the following sections.

In Section \ref{sec:t1flows} we consider the stationary $t_1$ flow in 2 coordinate systems, $\bf q$ (1 degree of freedom) and $\bf Q$ ($N$ degrees of freedom), deriving the specific case of the Hamiltonian (\ref{gen-t1flow}) for $M=3$. For each case we derive the Lax representation. When $N=1$, we derive a Poisson map, which gives a second (non-canonical) Poisson bracket for each of the systems.

A similar approach to the stationary $t_2$ flow is presented in Section \ref{sec:stat-t2}, where now the $\bf q$ space has 2 degrees of freedom, with the $\bf Q$ space still $N$ degrees of freedom.  When $N=2$, we derive a Poisson map, which gives a second (non-canonical) Poisson bracket for each of the systems.

Both of these stationary flows are actually {\em quadri-Hamiltonian} (for $M=3$), but to understand this we need to consider the Miura maps of Section \ref{sec:Miura}, but in the context of the stationary flows.  For $M=3$, we have 4 spaces, labelled $u, w, v$ and $z$ (in the PDE case), giving us 4 {\em stationary manifolds}, labelled $\bf q$, $\bf \hat Q$, $\bf \bar Q$ and $\bf \tilde Q$, which are extended to include 3 additional parameters as dynamical variables.  The latter are related to Casimirs of the Hamiltonian operators of the original PDEs.  These $\bf q$ coordinates coincide with our previous $\bf q$, but the others are different from $\bf Q$.  On the stationary manifolds, the Miura maps give rise to {\em Poisson maps}, leading to the construction of a sequences of 4 Poisson brackets for each of these spaces.  This construction is carried out for the $t_1$ and $t_2$ flows in Sections \ref{sec:quadHt1} and \ref{sec:quadHt2} respectively.  For the $\bf q$ space, the first two Poisson brackets coincide with the ones constructed in Sections \ref{sec:t1flows} and \ref{sec:stat-t2}.  We can then use the Poisson map between $\bf q$ and $\bf Q$ spaces to build a quadri-Hamiltonian formulation in the $\bf Q$ space.
In Section \ref{sec:quadHt2} we do this for the stationary $t_2$ flow (\ref{B1-t2-N=2}), which is separable in parabolic coordinates.  We analyse the structure of the Poisson tensors and, in Section \ref{sec:t2-recursion}, introduce a type of {\em recursion operator} connecting the core parts of the tensors.  The remaining part of the tensor is built in a specific way from the two commuting functions $h^{(Q)}$ and $f^{(Q)}$.

In Section \ref{sec:genPoisson}, we consider the general potential, separable in parabolic coordinates (see Equation 2.2.41 in \cite{90-16}).  We consider a corresponding generalisation of the 4 Poisson brackets found for the stationary $t_2$ flow (\ref{B1-t2-N=2}) in $\bf Q$ space.  Remarkably, the only restriction on the general potential is that they have a specific linear dependence on the Casimir parameters mentioned above.  However, imposing the Lax representation {\em does} restrict the potential to the specific forms related to the stationary reduction.  We show that it is possible to build the Lax matrices in a recursive way, avoiding the need to go through the lengthy process of the stationary reduction.


\section{Lax Representation of cKdV Equations}\label{sec:Lax-background}

In \cite{f87-5,f89-2} a detailed analysis was given of cKdV equations, associated with ``energy dependent'' Schr\"odinger operators.  In \cite{f23-1} we gave a very brief review, which we further reduce here.

\smallskip
This differential operator Lax pair for the $M$ component cKdV hierarchy is conveniently rewritten in ``zero-curvature'' form:
\begin{subequations}
\be\label{zero-c}
\left(
  \begin{array}{c}
    \psi_1 \\
    \psi_2
  \end{array}
\right)_x = \left(
              \begin{array}{cc}
                0 & 1 \\
                -u & 0 \\
              \end{array}
            \right) \left(
                      \begin{array}{c}
                        \psi_1 \\
                        \psi_2
                      \end{array}
                    \right)=U\Psi,\;\;\; \left(
                                     \begin{array}{c}
                                       \psi_1 \\
                                       \psi_2
                                     \end{array}
                                   \right)_{t} = \left(
                                                     \begin{array}{cc}
                                                       - P_x & 2 P \\[1mm]
                                                       -2 u P- P_{xx} &  P_x \\
                                                     \end{array}
                                                   \right)\left(
                                                            \begin{array}{c}
                                                              \psi_1 \\[1mm]
                                                              \psi_2
                                                            \end{array}
                                                          \right)=V \Psi ,
\ee
where $u = \sum_{i=0}^{M-1} u_i \lambda^i -\lambda^M$.  Consistency of this pair of equations leads to
\be\label{Ut}
U_{t}-V_x+[U,V]=0 \quad\Rightarrow\quad  u_t=(\pa_x^3+4 u \pa_x+2 u_x) P.
\ee
For a specific hierarchy, we must substitute the correct form of $u$ (by choosing $M$).

\smallskip
The cKdV hierarchy is given by the {\em polynomial expansion}
\be\label{Pol-P}
P^{(n)} = \sum_{k=0}^n \lambda^{n-k} P_k, \quad\mbox{with}\quad P_0=1,
\ee
corresponding to $t=t_n$.  The corresponding $V$ is called $V^{(n)}$, after which (\ref{zero-c}) takes the form
\be\label{Utn}
U_{t_n}-V^{(n)}_x+\left[U,V^{(n)}\right]=0, \;\;\;\mbox{with}\;\;\; u_{t_n}=(\pa_x^3+4 u \pa_x+2 u_x) P^{(n)}.
\ee
Equating coefficients of powers of $\lambda$ gives us a recursive formula for $P_k$ and then formulae for $u_{it_n}$, which can be written in matrix form to give the Hamiltonian formulation.
\end{subequations}

\subsection{The Hierarchy when $M=3$}

In \cite{f23-1} we considered the KdV ($M=1$) and DWW ($M=2$) cases.  Here we use the case $M=3$ to illustrate our results.

\smallskip
For the case $M=3$ we have $u=u_0+u_1\lambda + u_2 \lambda^2-\lambda^3$.  The coefficients $P_k$ in (\ref{Pol-P}) are constructed recursively, the first few being
\begin{subequations}  {\small
\bea
&& P_0 = 1,\quad P_1=\frac12u_2,\quad P_2=\frac{1}{8}(4u_1+3u_2^2),\quad P_3= \frac{1}{16}(8u_0+12u_1u_2+5u_2^3), \nn \\
&& P_4=\frac34u_0u_2+\frac38u_1^2+\frac{15}{16}u_1u_2^2+\frac{35}{128}u_2^4+\frac18u_{2xx}, \label{Pi-1to5} \\
&& P_5=\frac34u_0u_1+\frac{15}{16}u_2(u_0u_2+u_1^2)+\frac{35}{32}u_1u_2^3+\frac{63}{256}u_2^5+\frac18u_{1xx} +\frac5{16}u_2u_{2xx}+\frac5{32}u_{2x}^2.   \nn
\eea  }

The general recursion, defined by (\ref{Utn}), is
\be\label{JP=0}
J_0P_{n-3}+J_1P_{n-2}+J_2P_{n-1}+J_3P_n=0,
\ee
with\; $J_0= \pa_x^3+4 u_0 \pa_x+2 u_{0x},\;\; J_1= 4 u_1 \pa_x+2\, u_{1x},\;\; J_2= 4 u_2 \pa_x+2\, u_{2x},\;\; J_3 = -4\pa_x$.

\smallskip
For $P=P^{(n)}$, we have
\be\label{utn}
u_{0t_n}+\lambda u_{1t_n}+\lambda^2 u_{2t_n} = J_0P_n+\lambda (J_0P_{n-1}+J_1P_n)+\lambda^2(J_0P_{n-2}+J_1P_{n-1}+J_2P_n),
\ee
which can be written in vector form
\bea
\left(
  \begin{array}{c}
    u_0 \\
    u_1 \\
    u_2
  \end{array}
\right)_{t_n} &=& \left(
                  \begin{array}{ccc}
                    0 & 0 & J_0 \\
                    0 & J_0 & J_1 \\
                    J_0 & J_1 & J_2
                  \end{array}
                \right)\left(
                         \begin{array}{l}
                           P_{n-2} \\
                           P_{n-1} \\
                           P_n
                         \end{array}
                       \right)  =
                       \left(
                  \begin{array}{ccc}
                    0 & J_0 & 0\\
                    J_0 & J_1 & 0 \\
                    0 & 0 & -J_3
                  \end{array}
                \right)\left(
                         \begin{array}{l}
                           P_{n-1} \\
                           P_n\\
                           P_{n+1}
                         \end{array}
                       \right)    \nn  \\[-2mm]
                       &&  \label{ut=Bgrad}   \\[-2mm]
                       &=&
                       \left(
                  \begin{array}{ccc}
                    J_0 & 0 & 0 \\
                    0 & -J_2 & -J_3 \\
                    0 & -J_3 & 0
                  \end{array}
                \right)\left(
                         \begin{array}{l}
                           P_n\\
                           P_{n+1} \\
                           P_{n+2}
                         \end{array}
                       \right)
                       =\left(
                  \begin{array}{ccc}
                    -J_1 & -J_2 & -J_3 \\
                    -J_2 & -J_3 & 0 \\
                    -J_3 & 0 & 0
                  \end{array}
                \right)\left(
                         \begin{array}{l}
                           P_{n+1}\\
                           P_{n+2}\\
                           P_{n+3}
                         \end{array}
                       \right),    \nn
\eea
with the first being given directly by (\ref{utn}).  The general recursion (\ref{JP=0}) then gives the remaining representations.
\end{subequations}

These matrices define the 4 compatible Hamiltonian operators of this cKdV hierarchy, labelled respectively as $B_3,\ B_2,\, B_1$ and $B_0$.  Defining
\begin{subequations}\label{Hn}
\be\label{delHn}
\left(
                        \begin{array}{c}
                          P_{n-2}\\
                          P_{n-1} \\
                          P_n \\
                        \end{array}
                      \right) = \delta_u H_n = \left(
                                                 \begin{array}{c}
                                                   \delta_{u_0} H_n \\
                                                   \delta_{u_1} H_n \\
                                                   \delta_{u_2} H_n
                                                 \end{array}
                                               \right),
\ee
we have
\be\label{4Ham}
  {\bf u}_{t_n} = B_3 \delta_u H_n = B_2 \delta_u H_{n+1} = B_1 \delta_u H_{n+2}=B_0 \delta_u H_{n+3}, \quad\mbox{for}\;\;\; n\geq 0,
\ee
where ${\bf u} = (u_0,u_1,u_2)^T$ and     {\small
\bea
&& H_0 = u_2,\;\; H_1 = \frac{1}{4} (4u_1+ u_2^2),\;\; H_2 = u_0+\frac{1}{8} u_2 (4u_1+ u_2^2),\;\;  H_3 = \frac12u_0u_2+\frac{1}{64} (4 u_1+u_2^2)(4 u_1+5u_2^2),   \nn  \\
&& H_4=\frac{1}{8} u_0 (4u_1+3u_2^2)+ \frac{1}{128} u_2 (4 u_1+u_2^2)(12 u_1+7u_2^2)  -\frac1{16}u_{2x}^2,  \label{H1to5}   \\
&&  H_5=\frac{1}{16}u_0(4 u_0+12u_1u_2+5u_2^3)+\frac{1}{512}(4 u_1+u_2^2)(16 u_1^2+ 56 u_1u_2^2+21 u_2^4)-\frac{1}{32} u_{2x} (4 u_{1x}+5u_2u_{2x}).  \nn
\eea
}We also have
\be\label{BiCas}
B_0\delta_u H_0=B_0\delta_u H_1=B_0\delta_u H_2=B_1\delta_u H_0=B_1\delta_u H_1=B_2\delta_u H_0=0,
\ee
meaning that $B_0$ has {\em three local Casimir functions} $H_0,\, H_1,\, H_2$, whilst $B_1$ has {\em two local Casimir functions} $H_0,\ H_1$, with $B_2$ having just {\em one local Casimir function} $H_0$.
\end{subequations}

\br
Corresponding to these densities, we have an array of {\em fluxes} $F_{nm}$, in local conservation laws, given by the $t_n$ evolution of $H_m$:
$$
\pa_{t_n} H_m = \pa_x F_{nm}.
$$
In \cite{f23-1} we used these to build first integrals for stationary flows, but the same integrals arise from the Lax representation as well as the multi-Hamiltonian ladders, so we omit the formulae for $F_{nm}$ here.
\er

\subsubsection{Miura Maps}\label{sec:Miura}

In \cite{f89-2}, Miura maps were presented for the entire class of systems described by the ``energy dependent'' Schr\"odinger operator.  In the $M=3$ case, there are 4 sets of variables ${\bf u},\;{\bf w}=(w_0,w_1,w_2)^T,\; {\bf v}=(v_0,v_1,v_2)^T$ and ${\bf z}=(z_0,z_1,z_2)^T$, related by the following Miura maps
\begin{subequations}
\be\label{m1m2m3}
\begin{array}{llll}
  M_1[{\bf w}]: & u_0 = -w_{0x}-w_0^2, & u_1=w_1, & u_2=w_2,  \\
  M_2[{\bf v}]:  & w_0=v_0, & w_1 = -v_{1x}-2 v_0v_1, & w_2=v_2, \\
  M_3[{\bf z}]: & v_0 =z_0, & v_1=z_1, & v_2=-z_{2x}-2 z_0z_2-z_1^2.
  \end{array}
\ee
We have the 4 {\em local} Hamiltonian operators $B_i^u\equiv B_i$, given by (\ref{4Ham}), with 3 {\em local} operators $B_3^w$, $B_2^w$, $ B_1^w$ in the $w-$space, 2 {\em local} operator $B_3^v$, $B_2^v$ in the $v-$space and just 1 {\em local} operator $B_3^z$ in the $z-$space.  These are depicted in Figure \ref{B3B2B1B0-fig} and related by
\bea
&&  B_k^u = \frac{Du}{Dw}  B_k^w \left(\frac{Du}{Dw}\right)^\dagger\;\;\;\mbox{for}\;\;\; k=1,2,3, \nn\\[-2mm]
&&    \label{Bk-rels}   \\[-2mm]
&&  B_k^w = \frac{Dw}{Dv}  B_k^v \left(\frac{Dw}{Dv}\right)^\dagger\;\;\;\mbox{for}\;\;\; k=2,3 \qquad\mbox{and}\qquad   B_3^v = \frac{Dv}{Dz}  B_3^z \left(\frac{Dv}{Dz}\right)^\dagger ,  \nn
\eea
where $\frac{Du}{Dw}$, $\frac{Dw}{Dv}$ and $\frac{Dv}{Dz}$ are the Jacobians of the maps $M_1[{\bf w}]$, $M_2[{\bf v}]$ and $M_3[{\bf z}]$, respectively, given in (\ref{m1m2m3}), and the Hamiltonian operators in $w-$, $v-$ and $z-$space take the form
{\small
\bea
&&  B_3^w = \left(
\begin{array}{ccc}
0 &0 &-\pa_x^2+2\pa_x w_0\\
0 &\pa_x^3-4(w_0^2+w_{0x})\pa_x-2(2w_0w_{0x}+w_{0xx}) &4w_1\pa_x+2w_{1x}\\
\pa_x^2+2w_0\pa_x &4w_1\pa_x+2w_{1x} &4w_2\pa_x+2w_{2x}
\end{array}\right),    \nn  \\
&& B_2^w = \left(
\begin{array}{ccc}
0 &-\pa_x^2+2\pa_x w_0 &0\\
\pa_x^2+2w_0\pa_x &4w_1\pa_x+2w_{1x} &0\\
0 &0 &4\pa_x
\end{array}\right),
\quad
B_1^w =  \left(
\begin{array}{ccc}
-\pa_x &0 &0\\
0 &-4w_2\pa_x-2w_{2x} &4\pa_x\\
0 &4\pa_x &0
\end{array}\right), \label{Bk-forms} \\
&&   B_3^v = \left(
\begin{array}{ccc}
0 & 0 &-\pa_x^2\!+\!2\pa_x v_0\\
0 & -\pa_x &2\pa_x v_1\\
\pa_x^2\!+\!2v_0\pa_x &2v_1\pa_x &4v_2\pa_x\!+\!2v_{2x}
\end{array}\right),
\;\;
B_2^v = \left(
\begin{array}{ccc}
0 &-\pa_x  &0\\
-\pa_x &0  &0\\
0 &0 &4\pa_x
\end{array}\right),
\;\;
B_3^z =  \left(
\begin{array}{ccc}
0 &0 &-\pa_x\\
0 &-\pa_x &0\\
-\pa_x &0 &0
\end{array}\right).   \nn
\eea
}
\end{subequations}

{\small
\begin{figure}[hbt]
\begin{center}
\caption{Hamiltonian Operators in the 4 Spaces} \label{B3B2B1B0-fig}\vspace{6mm}
\unitlength=0.5mm
\begin{picture}(80,60)
\put(-20,60){\makebox(0,0){$B_3^u$}}
\put(-20,40){\makebox(0,0){$B_2^u$}}
\put(-20,20){\makebox(0,0){$B_1^u$}}
\put(-20,0){\makebox(0,0){$B_0^u$}}
\put(20,60){\makebox(0,0){$B_3^w$}}
\put(20,40){\makebox(0,0){$B_2^w$}}
\put(20,20){\makebox(0,0){$B_1^w$}}
\put(60,60){\makebox(0,0){$B_3^v$}}
\put(60,40){\makebox(0,0){$B_2^v$}}
\put(100,60){\makebox(0,0){$B_3^z$}}
\put(10,60){\vector(-1,0){20}}
\put(10,40){\vector(-1,0){20}}
\put(10,20){\vector(-1,0){20}}
\put(50,60){\vector(-1,0){20}}
\put(50,40){\vector(-1,0){20}}
\put(90,60){\vector(-1,0){20}}
\end{picture}
\end{center}
\end{figure}
}

We have a sequence of Hamiltonians $H_k^u\equiv H_k$, given by (\ref{Hn}), as functions of $u_i$ and their $x-$derivatives.
The Miura maps then define $H_k^w$, $H_k^v$ and $H_k^z$, by substituting the formulae \eqref{m1m2m3} into $H_k^u$.
The flow ${\bf u}_{t_n}$, defined by (\ref{4Ham}), gives rise to the flows:
\bea
 && {\bf w}_{t_n} =B_3^w \delta_w H_n^w = B_2^w \delta_w H_{n+1}^w=B_1^w \delta_w H_{n+2}^w,  \nn\\[-2mm]
  &&  \label{wtn-vtn}   \\[-2mm]
  &&    {\bf v}_{t_n} =B_3^v \delta_v H_n^v=B_2^v \delta_v H_{n+1}^v   \qquad\mbox{and}\qquad {\bf z}_{t_n} =B_3^z \delta_z H_n^z.    \nn
\eea

\subsection{The Stationary Flows}\label{sec:stat}

A {\em stationary} flow of (\ref{4Ham}) (for $t_n$) means that we reduce to a {\em finite dimensional} space with ${\bf u}_{t_n}=0$. The ``time'' variable for this system is $x$, which is the variable which appears in the Lagrangians,
given below.
\begin{enumerate}
  \item Using $B_0$ and in view of $B_0\delta_u H_0=B_0\delta_u H_1=B_0\delta_u H_2=0$, we have
\begin{subequations}
\be
B_0 \delta_u H_{n+3}=0 \quad\Rightarrow\quad    \delta_{u} (H_{n+3}-\alpha_0 H_0-4\alpha_1H_1-8\alpha_2H_2)=0,
\ee
which gives the first Lagrangian:
\be\label{Lag:B0}
{\cal L}_{n+3} = H_{n+3}-\alpha_0u_2-\alpha_1(4u_1+u_2^2)-\alpha_2(8u_0+4u_1u_2+u_2^3).
\ee
\end{subequations}
We then use the (generalised) Legendre transformation to find canonical coordinates $(q_i,p_i)$ and the Hamiltonian function.
  \item Using $B_1$ and the squared eigenfunction representation (following \cite{f95-3}), we have
   \begin{subequations}\label{B1-squared}
  \be\label{dH=phi2}
\delta_{u_0} H_{n+2} = -\frac{1}{2} \varphi^2 \quad\Rightarrow\quad  \varphi^2 (J_0 \varphi^2) = \left(2 \varphi^3 (\varphi_{xx} +u_0 \varphi) \right)_x = 0
                        \quad\Rightarrow\quad  2 \varphi^3 (\varphi_{xx} +u_0 \varphi) = 4 \beta,
\ee
and
\be
\delta_{u_1}H_{n+2}=4b_1,\quad \delta_{u_2}H_{n+2}=2b_1 u_2+b_0.
\ee
The exact derivative of \eqref{dH=phi2} arises as a consequence of the {\em skew symmetry} of $J_0$.
We can write the above equations as variational derivatives of a single Lagrangian function:
\be\label{Lag:B1}
{\cal L}_{n+2} = H_{n+2}-b_0 u_2-b_1(4u_1+u_2^2)+ \frac12 u_0 \varphi^2-\frac12\, \varphi_x^2+\frac{\beta}{\varphi^2},
\ee
with the above equations being given by $\delta_{u} {\cal L}_{n+2} =0,\;\;\delta_\varphi {\cal L}_{n+2}=0$.
\end{subequations}

Again, the (generalised) Legendre transformation gives canonical coordinates $(Q_i,P_i)$ and the Hamiltonian function.
\end{enumerate}

\br
The Hamiltonians listed in (\ref{Hn}) are built algorithmically, so often have inconvenient coefficients.  For our stationary flows, we typically choose a constant multiple of these in the formulae (\ref{Lag:B0}) and (\ref{Lag:B1}).
\er

Since $J_0$ is a linear operator we can extend the squared eigenfunction representation (\ref{dH=phi2}) to include multiple eigenfunctions (see \cite{f23-1} for more details), with
\begin{subequations}\label{B1-squaredN}
\be\label{mult-phi}
\delta_{u_0} H_{n+2}= -\frac{1}{2}\sum_i\varphi_i^2,
\ee
which yields \; $\delta_{u} {\cal L}_{n+2} =0,\;\;\; \delta_{\varphi_i} {\cal L}_{n+2}=0$, with
\be\label{Ln+2-phii-1}
{\cal L}_{n+2} = H_{n+2}-b_0 u_2-b_1(4u_1+u_2^2)+\sum_i \left(\frac{1}{2} u_0 \varphi_i^2-\frac{1}{2}\, \varphi_{ix}^2+\frac{\beta_i}{\varphi_i^2}\right),
\ee
or
\be\label{Ln+2-phii-2}
{\cal L}_{n+2} = H_{n+2}-b_0 u_2-b_1(4u_1+u_2^2)+\sum_i \left(\frac{1}{2} u_0 \varphi_i^2-\frac{1}{2}\, \varphi_{ix}^2\right)+\frac{\beta}{{\bf\varphi}^2},
\ee
where $\varphi^2 = \sum_i \varphi_i^2$.
\end{subequations}

\br
The Lagrangian in \eqref{Ln+2-phii-2} is rotationally invariant in the $\varphi$ space, so the {\em angular momenta} $(\varphi_i \varphi_{jx}-\varphi_j \varphi_{ix})$ are {\em constants of the motion}.  These can be generalised for the case of (\ref{Ln+2-phii-1}).
\er

\subsubsection{The Lax Representation of a Stationary Flow}

Setting $U_{t_n}=0$ in (\ref{Utn}), we obtain
\be\label{Vnx}
V^{(n)}_x = \left[U,V^{(n)}\right],
\ee
with these matrices being written in terms of $u_i$ and their $x-$derivatives.  The above Legendre transformations give us formulae for these variables in terms of the canonical coordinates, in terms of which (\ref{Vnx}) gives a Lax representation for the stationary flow, with $V^{(n)}$ playing the role of the Lax matrix $L$.


\section{The Stationary $t_1$ Flows}\label{sec:t1flows}

We first build the Lagrangians and Hamiltonians.  We use the corresponding Lax pair to construct the first integrals.  A bi-Hamiltonian representation is then constructed.

\subsection{Lagrangians and Legendre Transformations}

We build 3 different Hamiltonians.

\paragraph{The Lagrangian \eqref{Lag:B0},} using $8 H_4$ and labelling $-{\cal L}_4$ as $\cal L$, gives
\begin{subequations}  {\small
\be\label{B0-t1}
{\cal L} = \frac{1}{2} u_{2x}^2-4u_0u_1-3 u_2 \left(u_0u_2+u_1^2\right)-\frac{1}{16} u_2^3 \left(40 u_1+ 7 u_2^2\right) +\alpha_0u_2+\alpha_1(4u_1+u_2^2)+\alpha_2(8u_0+4u_1u_2+u_2^3).
\ee
}This Lagrangian is degenerate, but
$$
\delta_{u_0} {\cal L}=\delta_{u_1} {\cal L}=0 \quad\Rightarrow\quad  u_0=\alpha_1-2\alpha_2u_2+\frac12 u_2^3,\;\;\; u_1=2\alpha_2-\frac34u_2^2,
$$
after which\; ${\cal L} = \frac12u_{2x}^2+(\alpha_0-4\alpha_2^2)u_2-2\alpha_1u_2^2+2\alpha_2u_2^3-\frac14u_2^5+8\alpha_1\alpha_2$,\; which gives the Hamiltonian
\be\label{B0-t1-hq}
h^{(q)} =\frac12p_1^2+(4\alpha_2^2-\alpha_0)q_1+2\alpha_1q_1^2-2\alpha_2q_1^3+\frac14q_1^5,
\ee
where $q_1=u_2,\, p_1=u_{2x}$.
\end{subequations}

\paragraph{The Lagrangian \eqref{Ln+2-phii-1},} using $H_3$ and labelling $-{\cal L}_3$ as $\tilde{\cal L}$, gives
\begin{subequations}\label{B1-t1-squared}
\be\label{B1-t1}
\tilde {\cal L} = \sum_{i=1}^N \left(\frac{1}{2} \varphi_{ix}^2 -\frac{1}{2} u_0 \varphi_i^2-\frac{\beta_i}{\varphi_i^2}\right)
                   +b_0u_2+b_1(4u_1+u_2^2)- \frac{1}{2} u_0u_2  - \frac{1}{64} (4 u_1+u_2^2)(4 u_1+5 u_2^2).
\ee
This is also degenerate, but $\delta_{u_i}\tilde {\cal L}=0$, with $i=0,1,2$, give
$$
u_0=8b_1 \varphi^2-\frac12 \varphi^6+2 b_0,\;\; u_1=-\frac34 \varphi^4+8 b_1, \;\; u_2=-\varphi^2,\qquad \mbox{where} \quad \varphi^{2n} = \left(\sum_{i=1}^N \varphi_i^2\right)^n,
$$
leading to
\be\label{B1-t1-1}
\tilde {\cal L} = \sum_{i=1}^N \left(\frac12\varphi_{ix}^2 -b_0 \varphi_i^2-\frac{\beta_i}{\varphi_i^2}\right)
                              -2b_1 \varphi^{4}+\frac1{16}\varphi^{8}+16b_1^2.
\ee
Defining $Q_i=\varphi_i,\, P_i = \varphi_{ix}$ and removing the additive constant, we obtain the Hamiltonian
\be\label{B1-t1-hQ}
h^{(Q)} =\sum_{i=1}^N \left(\frac12P_i^2+b_0Q_i^2+\frac{\beta_i}{Q_i^2}\right)+2b_1{\bf Q}^4-\frac1{16}{\bf Q}^8, \qquad\mbox{where}\quad {\bf Q}^{2n}=\left(\sum_{i=1}^N Q_i^2\right)^n.
\ee
\end{subequations}

\paragraph{The Lagrangian \eqref{Ln+2-phii-2}}gives (removing an additive constant)
\begin{subequations}
\be\label{B1-t1-hQ-rot}
h^{(Q)} =\sum_{i=1}^N \left(\frac12P_i^2+b_0Q_i^2\right)+2b_1{\bf Q}^4-\frac1{16}{\bf Q}^8+\frac{\beta}{{\bf Q}^2}.
\ee

\smallskip
Clearly, this Hamiltonian is rotationally invariant, so the angular momenta are first integrals.
\end{subequations}

\subsection{The Lax Representation}\label{sec:stat-t1-Lax}

We now use the stationary Lax representation, given by (\ref{Vnx}), with the specific form $P^{(1)}=\lambda+\frac{1}{2} u_2$, defined by (\ref{Pol-P}), with (\ref{Pi-1to5}).  Specifically, for $M=3$, we have
\begin{subequations}
\bea
&&  U= \left(\begin{array}{cc}
           0 & 1 \\
           \lambda^3-u_2\lambda^2-u_1\lambda-u_0 & 0
           \end{array}\right), \qquad
V^{(1)}=\frac12 \left(\begin{array}{cc}
                     -u_{2x} & 4\lambda+2u_2 \\
                     a_{21} &  u_{2x}
                     \end{array}\right),  \nn\\[-2mm]
                     &&  \label{Lax-UV1}    \\[-2mm]
   \mbox{where} && a_{21}=2(\lambda^3-u_2\lambda^2-u_1\lambda-u_0)(2\lambda+u_2)-u_{2xx}.   \nn
\eea
\end{subequations}

\paragraph{Using the coordinates of the Hamiltonian (\ref{B0-t1-hq}),} we define $L^{(1)} = -2 V^{(1)}$, with
\begin{subequations}
\be\label{L1q}
\left.  \begin{array}{l}
          u_0=\alpha_1-2 \alpha_2 q_1+\frac{1}{2} q_1^3, \\
          u_1=2 \alpha_2-\frac{3}{4} q_1^2,\; u_2=q_1, \\
          u_{2xx}=q_{1xx}=-\frac{\pa h^{(q)}}{\pa q_1}
        \end{array}\right\}  \quad\Rightarrow\quad
L^{(1)} = \left(\begin{array}{cc}
               p_1 & -4\lambda-2q_1 \\
               a_{21}& -p_1
               \end{array}\right),
\ee
where $a_{21}$ is given by (\ref{Lax-UV1}) after substituting canonical coordinates.

\smallskip\noindent
The characteristic equation of $L^{(1)}$ is
\be\label{L1q-char}
 z^2-16\lambda^5+32\alpha_2\lambda^3+16\alpha_1\lambda^2 +4(\alpha_0-4\alpha_2^2)\lambda-2 h^{(q)}=0,
\ee
with $h^{(q)}$ given by (\ref{B0-t1-hq}).

\smallskip\noindent
The Lax representation of the equation generated by $h^{(q)}$ is
\be\label{L1qx}
L^{(1)}_x = \{L^{(1)},h^{(q)}\} = [U,L^{(1)}],
\ee
with $U$ given by (\ref{Lax-UV1}), after substituting for $u_i$.
\end{subequations}

\paragraph{Using the coordinates of the Hamiltonian (\ref{B1-t1-hQ}),} we define $L^{(1)} = V^{(1)}$ to obtain
\begin{subequations}
\be\label{L1-phi}
\left.  \begin{array}{l}
          u_0= 8 b_1 {\bf Q}^2-\frac{1}{2} {\bf Q}^6 +2 b_0, \\
          u_1= 8 b_1-\frac{3}{4} {\bf Q}^4,\; u_2= - {\bf Q}^2 \\
          u_{2xx}= -2 \sum_{i=1}^N \left(P_i^2-Q_i\frac{\pa h^{(Q)}}{\pa Q_i}\right)
        \end{array}\right\}  \quad\Rightarrow\quad
L^{(1)}= \left(\begin{array}{cc}
\sum_{1}^N Q_iP_i &  2\lambda -{\bf Q}^2 \\
a_{21} & -\sum_{1}^N Q_iP_i
\end{array}\right)  ,
\ee
with  \; $a_{21} = \frac{1}{4} \lambda (2 \lambda +{\bf Q}^2) (4 \lambda^2  + {\bf Q}^4 -32 b_1) -4 b_0 \lambda + \sum_{1}^N \left(P_i^2+\frac{2\beta_i}{Q_i^2}\right)$.

\smallskip\noindent
The characteristic equation of $L^{(1)}$ is
\be\label{L1Q-char}
z^2-4\lambda^5+32b_1\lambda^3+8b_0\lambda^2-4h^{(Q)}\lambda+2\Phi^{(Q)}=0,
\ee
where $h^{(Q)}$ is given by (\ref{B1-t1-hQ}) and
\be
 \Phi^{(Q)}=\frac{1}{2}\sum_{1\leq i<j\leq N} h_{ij}+\sum_{i=1}^N \beta_i,\quad\mbox{where}\quad h_{ij} = (Q_iP_j-Q_jP_i)^2 +2 \left(\beta_i\frac{Q_j^2}{Q_i^2}+\beta_j\frac{Q_i^2}{Q_j^2}\right).  \label{B1-t1-hij}
\ee
Indeed, each $h_{ij}$ is a first integral and, being deformations of the (squares of) angular momenta, it is straightforward to build an involutive set of integrals, so (\ref{B1-t1-hQ}) defines a {\em superintegrable system} (for $N\geq 3$).

\smallskip
This is a multi-component version of Case 1, Table II of \cite{90-22}, but with a particular form of potential.  However, this particular case has a Lax representation.

\smallskip
When $N=1$, $\Phi^{(Q)} = \beta_1$. When $N=2$, we have 2 integrals $h^{(Q)},\, h_{12}$.

\smallskip
When $N=3$, we have 4 integrals $h^{(Q)},\, h_{ij}$, with $h^{(Q)},\, h_{12}$ and $\Phi^{(Q)}$ in involution.

\smallskip
The Lax equations generated by $h^{(Q)}$ and $\Phi^{(Q)}$ are
\be\label{L1x-phi}
L^{(1)}_x = \{L^{(1)},h^{(Q)}\} = [U,L^{(1)}] , \quad \{L^{(1)}, \Phi^{(Q)}\} = 0,  \quad\mbox{with}\quad
U = \left(\begin{array}{cc}
 0 & 1 \\
 b_{21} & 0
\end{array}\right),
\ee
where \; $b_{21} = \frac{1}{2} (\lambda+{\bf Q}^2) (2 \lambda^2+{\bf Q}^4-16b_1)+\frac{1}{4}\lambda {\bf Q}^4-2 b_0$.
\end{subequations}

\smallskip\noindent
In fact, $\{L^{(1)}, h_{ij}\} = 0$, for each of the functions $h_{ij}$ of (\ref{B1-t1-hij}).

\br
It is known (see \cite{09-10}) that the general Hamiltonian (\ref{gen-t1flow}) is superintegrable, but {\em not} maximal, having $2N-3$ additional independent integrals.
However, the specific potentials derived as stationary reductions of the cKdV hierarchy also have a Lax representation.
\er

\paragraph{Using the coordinates of \eqref{B1-t1-hQ-rot},} we have the Lax representation (\ref{L1-phi})
with $\sum_{1}^N \frac{\beta_i}{Q_i^2}$ replaced by $\frac{\beta}{{\bf Q}^2}$ and the integral $\Phi^{(Q)}$ reduces to $\frac{1}{2}\sum_{i<j} (Q_iP_j-Q_jP_i)^2 + \beta$, reflecting the rotational symmetry.

\subsection{Bi-Hamiltonian Formulation when $N=1$}\label{sec:biHam-t1}

Here we have
\be\label{B1-t1-N=1}
h^{(Q)}=\frac12P_1^2+b_0Q_1^2+2b_1Q_1^4+\frac{\beta_1}{Q_1^2}-\frac1{16}Q_1^8.
\ee
The Hamiltonians (\ref{B0-t1-hq}) and (\ref{B1-t1-N=1}) define coordinates ${\bf q}=(q_1,p_1,\alpha_0,\alpha_1,\alpha_2)$ and ${\bf Q}=(Q_1,P_1,\beta_1,b_0,b_1)$, respectively.  From the definitions of the canonical coordinates, we have the following Poisson map
\be\label{PBmap-t1}
q_1=-Q_1^2,\quad p_1=-2Q_1P_1,\quad \alpha_0=-4 h^{(Q)}+64b_1^2,\quad \alpha_1=2b_0,\quad \alpha_2=4b_1,
\ee
with $h^{(q)} = -4 \beta_1$.

\br
The formulae for $\alpha_i$ and $h^{(q)}$ (in terms of the variables ${\bf Q}$) can be read directly from the formulae for the coefficients of the characteristic equations of the Lax matrices in the respective coordinates.
\er

We then follow the standard approach (outlined in Section 3.1.1 of \cite{f23-1}) to build a bi-Hamiltonian representation.

\smallskip
In the 5D spaces with coordinates $\bf q$ and $\bf Q$, we introduce the degenerate extension of the canonical Poisson tensor:
\begin{subequations}
\be\label{P00-P11-t1}
{\cal P}_0^{(q)} ={\cal P}_1^{(Q)}=
\left(\begin{array}{rrrrr}
0 & 1 & 0 & 0 & 0\\
-1 & 0 & 0 & 0 & 0 \\
0 & 0 & 0 & 0 & 0 \\
0 & 0 & 0 & 0 & 0 \\
0 & 0 & 0 & 0 & 0
\end{array}\right),
\quad\mbox{with}\;\;\; \frac{d{\bf q}}{dt_f} = {\cal P}_0^{(q)} \nabla_q f\ \mbox{and}\ \frac{d{\bf Q}}{dt_F} = {\cal P}_1^{(Q)} \nabla_Q F,
\ee
for any function $f(\bf{q})$ and $F(\bf{Q})$.  The formulae \eqref{PBmap-t1} represent a mapping from the $\bf Q$ space to the $\bf q$ space.  We use the Jacobian of the inverse map to construct a Poisson tensor ${\cal P}_0^{(Q)} = -4 \frac{\pa {\bf Q}}{\pa {\bf q}}\,  {\cal P}_0^{(q)}\, \left(\frac{\pa {\bf Q}}{\pa {\bf q}}\right)^T$ in the $\bf Q$ space, whose explicit form is given by
\be\label{P0Q-t1}
{\cal P}_0^{(Q)}=
\left(\begin{array}{ccccc}
 0 & -\frac{1}{Q_1^2}   & a_{13} &0 &0 \\
\frac{1}{Q_1^2} & 0 & a_{23}   &0 &0  \\
-a_{13}  &-a_{23}  &0 &0 &0 \\
0  &0 &0 &0 &0\\
0  &0 &0 &0 &0
\end{array}\right),
\ee
where $(a_{13},a_{23},0,0,0)^T={\cal P}_1^{(Q)}\nabla_Q h^{(Q)}$. This is compatible with the canonical bracket ${\cal P}_1^{(Q)}$ on the ${\bf Q}$ space,
which in turn allows us to construct a Poisson tensor \; ${\cal P}_1^{(q)}= -\frac{1}{4}\,\frac{\pa {\bf q}}{\pa {\bf Q}}\, {\cal P}_1^{(Q)} \left(\frac{\pa {\bf q}}{\pa {\bf Q}}\right)^T$, where
\be\label{P10-t1}
{\cal P}_1^{(q)} = \left(\begin{array}{ccccc}
0 & q_1 & b_{13} &0 &0 \\
-q_1 & 0 & b_{23} &0 &0  \\
-b_{13} &-b_{23} &0 &0 &0\\
0 &0 &0 &0 &0\\
0 &0 &0 &0 &0
\end{array}\right),
\ee
\end{subequations}
with $(b_{13},b_{23},0,0,0)^T={\cal P}_0^{(q)}\nabla_q h^{(q)}$.

\smallskip
Each of these Poisson matrices has three Casimirs and the $t_1$ ($=t_h$) flow in each space has a bi-Hamiltonian representation:
\begin{subequations}
\bea
&&  {\bf q}_{t_h}={\cal P}_0^{(q)}\nabla_q h^{(q)} = {\cal P}_1^{(q)}\nabla_q \,\alpha_0,    \nn   \\[-2mm]
    &&   \label{t1-q-flows}    \\[-2mm]
&&  {\cal P}_0^{(q)}\nabla_q \alpha_0={\cal P}_0^{(q)}\nabla_q \alpha_1={\cal P}_0^{(q)}\nabla_q \alpha_2
={\cal P}_1^{(q)}\nabla_q h^{(q)}={\cal P}_1^{(q)}\nabla_q \alpha_1={\cal P}_1^{(q)}\nabla_q \alpha_2=0,  \nn
\eea
in the ${\bf q}$ space and
\bea
&&  {\bf Q}_{t_h} = {\cal P}_1^{(Q)}\nabla_Q h^{(Q)}= {\cal P}_0^{(Q)}\nabla_Q \beta_1,    \nn   \\[-2mm]
  &&     \label{t1-Q-flows}    \\[-2mm]
&&  {\cal P}_1^{(Q)}\nabla_Q b_0={\cal P}_1^{(Q)}\nabla_Q b_1={\cal P}_1^{(Q)}\nabla_Q \beta_1  ={\cal P}_0^{(Q)}\nabla_Q b_0={\cal P}_0^{(Q)}\nabla_Q b_1={\cal P}_0^{(Q)}\nabla_Q h^{(Q)}=0,   \nn
\eea
in the ${\bf Q}$ space.
\end{subequations}


\section{The Stationary $t_2$ Flows}\label{sec:stat-t2}

We first build the Lagrangians and Hamiltonians.  We use the corresponding Lax pair to construct the first integrals.  A bi-Hamiltonian representation is then constructed.

\subsection{Lagrangians and Legendre Transformations}

We build 3 different Hamiltonians.

\paragraph{The Lagrangian \eqref{Lag:B0},} using $2 H_5$ and labelling ${\cal L}_5$ as $\cal L$, gives
\begin{subequations}
\bea
{\cal L} &=& \frac{1}{2} u_0^2 +\frac{1}{8} u_0 u_2 \left(12 u_1+5 u_2^2\right) +\frac{1}{256} \left(4 u_1+u_2^2\right)\left(16 u_1^2+56 u_1 u_2^2+21 u_2^4\right) -\frac{1}{4} u_{1x}u_{2x} \nn\\
&&  \hspace{4cm}     -\frac{5}{16} u_2u_{2x}^2 -\alpha_0u_2-\alpha_1(4u_1+u_2^2)-\alpha_2(8u_0+4u_1u_2+u_2^3).    \label{B0-t2}
\eea
This Lagrangian is degenerate, but $\delta_{u_0} {\cal L}=0$ gives $u_0=8\alpha_2-\frac32u_1u_2-\frac58 u_2^3$, after which
\bea
{\cal L} &=& \frac14u_1^3-\frac3{16}u_1^2u_2^2 -\frac{25}{64}u_1u_2^4 - \frac{29}{256}u_2^6 -\frac{1}{4} u_{1x}u_{2x}  -\frac{5}{16} u_2u_{2x}^2 \nn\\
        && \hspace{3cm}         -\alpha_0u_2-\alpha_1(4u_1+u_2^2) +4 \alpha_2 u_2 \left(2 u_1+u_2^2\right) -32 \alpha_2^2  ,
\eea
leading to $q_1=u_1,\; q_2=u_2,\; p_1=-\frac14u_{2x},\; p_2=-\frac58u_2u_{2x}-\frac14u_{1x}$ and (removing the additive constant) the Hamiltonian
\be\label{B0-t2-hq}
h^{(q)} =5q_2p_1^2-4p_1p_2+\alpha_0q_2+\alpha_1(4q_1+q_2^2)-4\alpha_2 q_2 (2q_1+q_2^2)-\frac14q_1^3+\frac3{16}q_1^2q_2^2  +\frac{25}{64}q_1q_2^4+\frac{29}{256}q_2^6.
\ee
\end{subequations}

\paragraph{The Lagrangian \eqref{Ln+2-phii-1},} using $8H_4$ and labelling $-{\cal L}_4$ as $\tilde{\cal L}$, we find
\begin{subequations}
{\small  \be\label{B1-t2}
\tilde {\cal L} = \sum_{i=1}^{N-1} \left(\frac{1}{2} \varphi_{ix}^2 -\frac{1}{2} u_0 \varphi_i^2-\frac{\beta_{i+1}}{\varphi_i^2}\right)+\frac{1}{2}u_{2x}^2
                 -\frac{u_2}{16}  \left(4 u_1+u_2^2\right)\left(12 u_1+7 u_2^2\right) -u_0\left(4 u_1+3 u_2^2\right) +b_0 u_2+b_1 (4u_1+u_2^2) .
\ee
}This is also degenerate, but $\delta_{u_i}\tilde {\cal L}=0$ with $i=0,1$ give \, $u_0=\frac3{16}u_2\sum_1^{N-1} \varphi_i^2+\frac{1}{2}u_2^3+b_1, \, u_1=-\frac{1}{8}\sum_1^{N-1} \varphi_i^2-\frac34u_2^2$, leading to
\be\label{B1-t2-1}
\tilde {\cal L} = \sum_{i=1}^{N-1}  \left(\frac12\varphi_{ix}^2 -\frac14(u_2^3+2b_1) \varphi_i^2-\frac{\beta_{i+1}}{\varphi_i^2}\right) +\frac12u_{2x}^2
                                     -\frac3{64}u_2\left(\sum_{i=1}^{N-1} \varphi_i^2\right)^2-\frac14u_2(u_2^4+8b_1u_2-4b_0).
\ee
Defining \; $Q_1=u_2,\, Q_{i+1}=\varphi_i,\,P _1=u_{2x},\, P_{i+1}=\varphi_{ix}$, for $i=1,\dots,N-1$, leads to the Hamiltonian
\be\label{B1-t2-hQ}
h^{(Q)} = \frac{1}{2}P_1^2+ \frac{1}{2} \sum_{i=2}^N \left(P_i^2+\frac{2\beta_i}{Q_i^2}\right) -b_0 Q_1 + \frac{1}{2} b_1 (4 Q_1^2+{\bf Q}^2) +\frac{1}{64} Q_1 (4 Q_1^2+{\bf Q}^2)(4 Q_1^2+3{\bf Q}^2).
\ee
\end{subequations}
where ${\bf Q}^2=\sum_{i=2}^N Q_i^2$.

\paragraph{The Lagrangian \eqref{Ln+2-phii-2}} gives
\be\label{B1-t2-hQ-rot}
h^{(Q)} = \frac12\sum_{i=1}^N P_i^2 -b_0 Q_1 + \frac{1}{2} b_1 (4 Q_1^2+{\bf Q}^2) +\frac{1}{64} Q_1 (4 Q_1^2+{\bf Q}^2)(4 Q_1^2+3{\bf Q}^2)+\frac{\beta}{{\bf Q}^2}.
\ee

\smallskip
Clearly, this Hamiltonian is rotationally invariant in the $Q_2,\dots ,Q_N$ space, so the angular momenta in this space are first integrals.

\subsection{The Lax Representation}\label{sec:stat-t2-Lax}

Again, we use the stationary Lax representation, given by (\ref{Vnx}), with the specific form $P^{(2)} = \lambda^2+\frac{1}{2} u_2 \lambda +\frac{1}{8} (4 u_1+3 u_2^2)$, defined by (\ref{Pol-P}) and (\ref{Pi-1to5}), giving
\be\label{V2}
V^{(2)}=\frac14 \left(\begin{array}{cc}
            -(2\lambda+3u_2)u_{2x}-2u_{1x} & 8\lambda^2+4u_2\lambda+3u_2^2+4u_1 \\
            b_{21} &  (2\lambda+3u_2)u_{2x}+2u_{1x}
            \end{array}\right),
\ee
with \; $b_{21}=(\lambda^3-u_2\lambda^2-u_1\lambda-u_0)(8\lambda^2+4u_2\lambda+3u_2^2+4u_1) -  2\lambda u_{2xx} -(3 u_2 u_{2x}+2 u_{1x})_x$.

\smallskip
We also need $U$ and $V^{(1)}$ of (\ref{Lax-UV1}), but now written in terms of the stationary $t_2$ coordinates.

\paragraph{Using the coordinates of the Hamiltonian (\ref{B0-t2-hq}),} we define $L^{(2)}=\frac{1}{2} V^{(2)}$ to obtain
\begin{subequations}
\be\label{L2-q}
\left.  \begin{array}{l}
          u_0= 8 \alpha_2-\frac{3}{2} q_1q_2-\frac{5}{8} q_2^3, \\
          u_1= q_1,\; u_2= q_2,\; u_{2x}=-4 p_1 \\
          u_{1x}= 2 (5 q_2p_1-2 p_2)
        \end{array}\right\}  \quad\Rightarrow\quad
L^{(2)}=\left(\begin{array}{cc}
\lambda p_1 + p_2-q_2 p_1 & \lambda^2 +\frac{1}{2} q_2 \lambda+\frac{1}{8} (4q_1+3 q_2^2)\\
                   a_{21} & q_2 p_1-p_2 - \lambda p_1
\end{array}\right)
\ee
where
$$
a_{21} = \frac{1}{8} \left(\lambda^3-q_2 \lambda^2-q_1 \lambda+\frac{3}{2} q_1 q_2+\frac{5}{8} q_2^3-8 \alpha_2\right)\left(8 \lambda^2+4 q_2 \lambda +4 q_1+3 q_2^2\right)
                                                -\lambda \frac{\pa h^{(q)}}{\pa q_1}- \frac{\pa f^{(q)}}{\pa q_1}.
$$
Here $f^{(q)}$ is given by
\bea
f^{(q)} &=& \frac{1}{4} \left(4q_1- 5q_2^2\right)p_1^2+4q_2p_1p_2-2p_2^2+\frac{1}{4}\alpha_0\left(4 q_1+3 q_2^2\right) -\frac{1}{2}\alpha_1q_2\left(4q_1+3q_2^2\right)  \nn\\
&&     \hspace{1.5cm}  -\frac{1}{4}\alpha_2(4 q_1+q_2^2)(4 q_1+3 q_2^2)+\frac{1}{512} q_2 (4 q_1+q_2^2)(4 q_1+3 q_2^2) (12 q_1+7 q_2^2), \label{t2-fq}
\eea
which is a first integral, appearing in the characteristic equation of $L^{(2)}$:
$$
z^2 +\frac{1}{2} f^{(q)} + \frac{1}{2} h^{(q)} \lambda + \alpha_0 \lambda^2 +4 \alpha_1 \lambda^3 +8\alpha_2\lambda^4-\lambda^7=0.
$$

\br
The last two terms in $a_{21}$ follow from the formulae:
$$
u_{2xx}=\left\{-4 p_1,h^{(q)}\right\}=4 \frac{\pa h^{(q)}}{\pa q_1} \quad\mbox{and}\quad (3 u_2 u_{2x}+2 u_{1x})_x=8\left\{q_2p_1-p_2,h^{(q)}\right\}=8 \frac{\pa f^{(q)}}{\pa q_1}.
$$
\er

\smallskip
The Lax representation of the equations generated by $h^{(q)}$ and $f^{(q)}$ are
\be\label{L2x}
L^{(2)}_x = \{L^{(2)},h^{(q)}\} = [U,L^{(2)}] , \quad  L^{(2)}_{t_1} = \{L^{(2)},f^{(q)}\} = [L^{(1)},L^{(2)}] ,
\ee
where
\be\label{UL1q}
U =\begin{pmatrix}
  0 & 1 \\
 \lambda^3-q_2\lambda^2-q_1\lambda+\frac32q_1q_2+\frac58q_2^3-8 \alpha_2 & 0
\end{pmatrix},
\quad
L^{(1)} = \left(\begin{array}{cc}
              p_1 & \lambda+\frac{1}{2}\, q_2 \\
              c_{21} & -p_1 \\
\end{array}\right),
\ee
with\; $c_{21} = \left(\lambda^3-q_2 \lambda^2-q_1 \lambda+\frac{3}{2} q_1 q_2+\frac{5}{8} q_2^3-8 \alpha_2\right)\left(\lambda+\frac{1}{2} q_2\right)-\frac{\pa h^{(q)}}{\pa q_1}$.
\end{subequations}

\paragraph{Using the coordinates of \eqref{B1-t2-hQ},} we define \; $L^{(2)} = 8 V^{(2)}$ to obtain
\begin{subequations}
\be\label{L2-Q}
\left.  \begin{array}{l}
          u_0= \frac{1}{16} Q_1 (8 Q_1^2+3 {\bf Q}^2)+b_1, \\[1mm]
          u_1= -\frac{1}{8} (6 Q_1^2+{\bf Q}^2),\\
          u_2= Q_1,\;\; u_{2x}= P_1 \\
          u_{1x}= -\frac{1}{4} \left(6 Q_1P_1+\sum_{i=2}^N Q_iP_i\right)
        \end{array}\right\}  \quad\Rightarrow\quad
L^{(2)} = \left(\begin{array}{cc}
\sum_{2}^N Q_iP_i -4 \lambda P_1 &  16\lambda^2 +8 \lambda Q_1-{\bf Q}^2 \\
                           a_{21} & 4 \lambda P_1 -\sum_{2}^N Q_iP_i
\end{array}\right) ,
\ee
where  {\small
\be \label{L2-hQfQ-a21}
   a_{21} = \left(\lambda^3-Q_1 \lambda^2+\frac{1}{8}(6 Q_1^2+{\bf Q}^2)\lambda-\frac{1}{16}Q_1(8Q_1^2+3 {\bf Q}^2)-b_1\right) \left(16 \lambda^2+8 Q_1 \lambda-{\bf Q}^2\right)
                            + 4 \frac{\pa h^{(Q)}}{\pa Q_1} \lambda - \frac{\pa f^{(Q)}}{\pa Q_1},
\ee
}with ${\bf Q}^2= \sum_2^N Q_i^2$. The characteristic equation of $L^{(2)}$ is
\be\label{L2-char-Q}
z^2 +2\Phi^{(Q)} +8 f^{(Q)} \lambda -32 h^{(Q)} \lambda^2+64 b_0 \lambda^3+ 256 b_1 \lambda^4- 256\lambda^7=0,
\ee
where $\Phi^{(Q)}$ and $f^{(Q)}$ are given by
{\small
\bea
 \Phi^{(Q)} &=& \frac{1}{2}\sum_{2\leq i<j\leq N} h_{ij}+\sum_{i=2}^N \beta_i,   \label{t2-PhiQ}\\
f^{(Q)} &=& P_1 \, \sum_{i=1}^N Q_i P_i-2 Q_1 h^{(Q)}+\frac12Q_1^6 +\frac1{128}{\bf Q}^2(4 Q_1^2+ {\bf Q}^2)(20 Q_1^2+{\bf Q}^2) \nn\\
&&  \hspace{5cm} -\frac12b_0\left(4 Q_1^2+{\bf Q}^2\right)+2b_1Q_1\left(2Q_1^2+{\bf Q}^2\right),  \label{t2-fQ}
\eea
}where $h_{ij}$ , given by (\ref{B1-t1-hij}), are themselves first integrals.

\smallskip
The Lax equations generated by $h^{(Q)},\, f^{(Q)}$ and $h_{ij}$ (not just the {\em combination} $\Phi^{(Q)}$) are
\bea
&&  L^{(2)}_x = \{L^{(2)},h^{(Q)}\} = [U,L^{(2)}] , \qquad L^{(2)}_{t_1} = \{L^{(2)},f^{(Q)}\} = [L^{(1)},L^{(2)}] , \nn\\[-2mm]
&&    \label{L2x-phi}    \\[-2mm]
&&   \{L^{(2)}, h_{ij}\} = 0,\;\;\; 2\leq i<j\leq N ,    \nn
\eea
where   {\small
\be\label{L1QU}
L^{(1)} = \left(\begin{array}{cc}
              P_1 & -4\lambda -2 Q_1 \\
              b_{21} & -P_1
\end{array}\right),
\quad U = \left(\begin{array}{cc}
0 & 1 \\
\lambda^3-Q_1\lambda^2+\frac{1}{8}\left(6 Q_1^2+{\bf Q}^2\right)\lambda-\frac{1}{16}Q_1(8Q_1^2+3 {\bf Q}^2)-b_1 & 0
\end{array}\right),
\ee
}with \;\;
$
b_{21} =  \left(\lambda^3-Q_1 \lambda^2+\frac{1}{8}(6 Q_1^2+{\bf Q}^2)\lambda-\frac{1}{16}Q_1(8Q_1^2+3 {\bf Q}^2)-b_1\right) \left( -4\lambda -2 Q_1\right)-\frac{\pa h^{(Q)}}{\pa Q_1}.
$
\end{subequations}

\paragraph{Using the coordinates of \eqref{B1-t2-hQ-rot},} we have the Lax representation \eqref{L2-Q}
with $\sum_{2}^N \frac{\beta_i}{Q_i^2}$ replaced by $\frac{\beta}{{\bf Q}^2}$ and the integral (\ref{t2-PhiQ}) reduces to
\be\label{t2-PhiQ-rot}
\Phi^{(Q)} = \frac{1}{2}\sum_{2\leq i<j\leq N}(Q_iP_j-Q_jP_i)^2+\beta,
\ee
which is just the Casimir ($+\beta$) of the rotation algebra in the $Q_2,\dots ,Q_N$ space.

\subsection{Bi-Hamiltonian Formulation when $N=2$}\label{sec:biHam-t2}

Here we have the Hamiltonian of the form
\be\label{B1-t2-N=2}
h^{(Q)}=\frac{1}{2}(P_1^2+P_2^2)+\frac{\beta_2}{Q_2^2} -b_0 Q_1 +\frac{1}{2} b_1 (4 Q_1^2+Q_2^2) +\frac{1}{64} Q_1 (4 Q_1^2+Q_2^2)(4 Q_1^2+3 Q_2^2).
\ee
The Hamiltonians (\ref{B0-t2-hq}) and (\ref{B1-t2-N=2}) define coordinates ${\bf q}=(q_i,p_i,\alpha_0,\alpha_1,\alpha_2)$ and ${\bf Q}=(Q_i,P_i,\beta_2,b_0,b_1)$, respectively.  From the definitions of the canonical coordinates, we have the Poisson map
\bea
&&   q_1=-\frac{1}{8} (6 Q_1^2+Q_2^2),\quad q_2=Q_1,\;\; p_1=-\frac14P_1,\;\; p_2= \frac1{16}(Q_2P_2-4Q_1P_1),\nn\\[-2mm]
    &&  \label{PBmap-t2}  \\[-2mm]
    &&   \alpha_0=-\frac18 h^{(Q)},\;\; \alpha_1=\frac1{16}b_0,\;\; \alpha_2=\frac18 b_1, \quad\mbox{together with}\quad h^{(q)}=\frac{1}{16} f^{(Q)},\;\; f^{(q)}=\frac{1}{64} \beta_2.  \nn
\eea

\br
Again, the formulae for $\alpha_i$ and $h^{(q)}, f^{(q)}$ (in terms of the variables ${\bf Q}$) can be read directly from the formulae for the coefficients of the characteristic equations of the Lax matrices in the respective coordinates.
\er

\smallskip
Again we introduce an extended canonical bracket on each space, labelled ${\cal{P}}_0^{(q)}$ and ${\cal{P}}_1^{(Q)}$, and use (\ref{PBmap-t2}) as a Poisson map to obtain the Poisson tensor ${\cal{P}}_0^{(Q)}$ and ${\cal{P}}_1^{(q)}$:
\begin{subequations}
{\small \be\label{t2-P0q-P0Q}
  {\cal{P}}_0^{(q)}=\left(\begin{array}{rrrrrrr}
0&0&1&0&0&0&0  \\
0&0&0&1&0&0&0  \\
-1&0&0&0&0&0&0 \\
0&-1&0&0&0&0&0 \\
0&0&0&0&0&0&0\\
0&0&0&0&0&0&0\\
0&0&0&0&0&0&0
\end{array}\right)
\quad\Rightarrow\quad
{\cal{P}}_0^{(Q)} = \left(\begin{array}{ccccccc}
  0&0&0&\frac1{Q_2} &a_{15} &0 &0   \\[2mm]
  0&0&\frac1{Q_2}&-\frac{2Q_1}{Q_2^2} & a_{25} &0 &0  \\[2mm]
  0&-\frac1{Q_2}&0&\frac{P_2}{Q_2^2} & a_{35} &0 &0  \\[2mm]
  -\frac1{Q_2}&\frac{2Q_1}{Q_2^2}&-\frac{P_2}{Q_2^2}&0  & a_{45} &0 &0 \\[2mm]
   -a_{15} & -a_{25} &-a_{35} & -a_{45} &0 &0 &0  \\
  0&0&0&0&0&0&0\\
  0&0&0&0&0&0&0
\end{array}\right),
\ee
}where the column $(a_{15},a_{25},a_{35},a_{45},0,0,0)^T=-P_1^{(Q)}\nabla_Q f^{(Q)}$, with ${\cal{P}}_1^{(Q)}$ being the compatible canonical bracket, from which we build ${\cal{P}}_1^{(q)}$:
{\small\be\label{t2-P10}
{\cal{P}}_1^{(q)}=
\left(\begin{array}{ccccccc}
0 & 0 & 12q_2 & 15q_2^2+4q_1 & b_{15} & 0 &0 \\
0 &0  &-8   &-8q_2                & b_{25} & 0 &0 \\
-12q_2  &8  &0  &8p_1  & b_{35} & 0 &0  \\
-15 q_2^2-4q_1 & 8q_2 &-8p_1 &0 & b_{45} & 0 &0\\
-b_{15} &-b_{25} &-b_{35} &-b_{45} &0 &0 &0\\
0&0&0&0&0&0&0\\
0&0&0&0&0&0&0
\end{array}\right),
\ee
}in which the column $(b_{15},b_{25},b_{35},b_{45},0,0,0)^T=-4P_0^{(q)}\nabla_q h^{(q)}$.
\end{subequations}

\smallskip
In the ${\bf q}$ space, the bi-Hamiltonian ladder is
\begin{subequations}
\bea
&&  {\bf q}_{t_h} = {\cal P}_0^{(q)}\nabla_q h^{(q)} = {\cal P}_1^{(q)}\nabla_q \left(-\frac{1}{4}\alpha_0\right),\quad
                                {\bf q}_{t_f} = {\cal P}_0^{(q)}\nabla_q f^{(q)} = {\cal P}_1^{(q)}\nabla_q \left( -\frac{1}{8}h^{(q)}\right), \nn\\[-2mm]
&&    \label{t2-q-flows}    \\[-2mm]
& &  {\cal P}_0^{(q)}\nabla_q \alpha_0={\cal P}_0^{(q)}\nabla_q \alpha_1={\cal P}_0^{(q)}\nabla_q \alpha_2 ={\cal P}_1^{(q)}\nabla_q f^{(q)}={\cal P}_1^{(q)}\nabla_q \alpha_1={\cal P}_1^{(q)}\nabla_q \alpha_2=0,  \nn
\eea
and in $\bf{Q}$ space we have
\bea
&&  {\bf Q}_{t_h} = {\cal P}_1^{(Q)}\nabla_Q h^{(Q)}={\cal P}_0^{(Q)}\nabla_Q f^{(Q)},\quad   {\bf Q}_{t_f} = {\cal P}_1^{(Q)}\nabla_Q f^{(Q)}= {\cal P}_0^{(Q)}\nabla_Q (-\beta_2),   \nn\\[-2mm]
&&      \label{t2-Q-flows}      \\[-2mm]
&&  {\cal P}_1^{(Q)}\nabla_Q b_0={\cal P}_1^{(Q)}\nabla_Q b_1={\cal P}_1^{(Q)}\nabla_Q \beta_2 ={\cal P}_0^{(Q)}\nabla_Q b_0={\cal P}_0^{(Q)}\nabla_Q b_1={\cal P}_0^{(Q)}\nabla_Q h^{(Q)}=0.  \nn
\eea
\end{subequations}
Here $f^{(q)}$ and $f^{(Q)}$ are given in (\ref{t2-fq}) and (\ref{t2-fQ}) (with $N=2$) respectively.


\section{The Stationary $t_1$ Flow in Quadri-Hamiltonian Form}\label{sec:quadHt1}

We now use the Miura maps of Section \ref{sec:Miura} to construct a {\em quadri-Hamiltonian} formulation of the {\em stationary flows} corresponding to (\ref{4Ham}) and (\ref{wtn-vtn}).  In this way, Figure \ref{B3B2B1B0-fig} is extended to an array of 16 Poisson matrices ${\cal P}_k^{(m)},\; k,m=0,1,2,3$.  Here the index ``$m$'' refers to the particular space, with the correspondence 
$(0,1,2,3)=\left({\bf q},{\bf \hat Q},{\bf \bar Q},{\bf \tilde Q}\right)$.

\subsection{Defining the Coordinates}

The 4 Poisson matrices ${\cal P}_k^{(k)}$ can be {\em directly constructed} in canonical form, giving us coordinates $(q_i,p_i)$, $(\hat Q_i,\hat P_i)$, $(\bar Q_i,\bar P_i)$ and $(\tilde Q_i,\tilde P_i)$ respectively.

\paragraph{Defining the $u$ space, using $B_0^u$:}

We previously defined these canonical coordinates when deriving (\ref{B0-t1-hq}), to obtain
\be\label{B0-t1-hu}
h^{(q)} =\frac12p_1^2+(4\alpha_2^2-\alpha_0)q_1+2\alpha_1q_1^2-2\alpha_2q_1^3+\frac14q_1^5.
\ee

\paragraph{Defining the $w$ space, using $B_1^w$:}

The stationary $t_n$ flow (from (\ref{wtn-vtn})) is defined by
\begin{subequations}\label{B1wdH=0}
\be
B_1^w \delta_w H_{n+2}^w=0 \quad\Rightarrow\quad  \delta_w \left(H_{n+2}^w-\hat \beta_0 w_0-\hat \beta_1 w_2-\hat \beta_2\left(4w_1+w_2^2\right)\right)=0,
\ee
which gives
\be\label{Lag:B1w}
{\cal L}_{n+2}^w =H_{n+2}^w-\hat \beta_0 w_0-\hat \beta_1 w_2-\hat \beta_2\left(4w_1+w_2^2\right).
\ee
For the $t_1$ flow, we use $4H_3^w$, giving
\be\label{t1-L3w}
{\cal L}_{3}^w = -2(w_{0x}+w_0^2)w_2+w_1^2+\frac32w_1w_2^2+\frac5{16}w_2^4-\hat \beta_0w_0-\hat \beta_1w_2-\hat \beta_2(4w_1+w_2^2).
\ee
This is degenerate and leads to $w_0=\frac{2w_{2x}-\hat \beta_0}{4 w_2}$ and $w_1=-\frac34w_2^2+2\hat \beta_2$.  Removing an exact derivative, we find
\be\label{t1-L3-1}
{\cal L}_{3}^w = \frac{(w_{2x}-\frac12\hat \beta_0)^2}{2 w_2}+\frac14w_2(8\hat \beta_2w_2-w_2^3-4\hat \beta_1),
\ee
corresponding to the Hamiltonian
\be\label{B1-t1-hw}
h^{(\hat Q)} = \frac{1}{2}\hat Q_1\hat P_1^2+\frac12\hat \beta_0 \hat P_1+\frac{1}{4} \hat Q_1 (\hat Q_1^3-8\hat \beta_2\hat Q_1+4\hat \beta_1),
\ee
where $\hat Q_1 = w_2, \, \hat P_1 = \frac{w_{2x}-\frac{1}{2} \hat\beta_0}{w_2}$.
\end{subequations}

\paragraph{Defining the $v$ space, using $B_2^v$:}

The stationary $t_n$ flow (from (\ref{wtn-vtn})) is defined by
\begin{subequations}
\be\label{B2vdH=0}
B_2^v \delta_v H_{n+1}^v=0 \;\; \Rightarrow\;\;  \delta_v \left(H_{n+1}^v-\gamma_0 v_0-\gamma_1 v_1-\gamma_2 v_2\right)=0,
\ee
which gives
\be\label{Lag:B2v}
{\cal L}_{n+1}^v = H_{n+1}^v-\gamma_0 v_0-\gamma_1 v_1-\gamma_2 v_2.
\ee
For the $t_1$ flow, we use $-H_2^v$, removing an exact derivative, to obtain
\be\label{t1-L2v}
{\cal L}_{2}^v =\frac12v_{1x}v_2+v_0^2+v_0v_1v_2-\frac18v_2^3-\gamma_0 v_0-\gamma_1 v_1-\gamma_2 v_2.
\ee
This is degenerate and leads to $v_0=\frac{v_{2x}+2\gamma_1}{2v_2}$ and $v_1=\frac{\gamma_0v_2-v_{2x}-2\gamma_1}{v_2^2}$.  Removing an exact derivative, we find
\be\label{t1-L2v-1}
{\cal L}_{2}^v =\frac{(v_{2x}-\gamma_0v_2)^2}{4v_2^2}-\frac{v_2^3}8+\frac{\gamma_1^2}{v_2^2}-\frac{\gamma_0\gamma_1}{v_2} -\gamma_2v_2,
\ee
which gives the Hamiltonian
\be\label{B2-t1-hv}
h^{(\bar Q)} =\bar Q_1^2\bar P_1^2+ \gamma_0 \bar Q_1 \bar P_1+\frac{\bar Q_1^3}8+\frac{\gamma_0\gamma_1}{\bar Q_1} -\frac{\gamma_1^2}{\bar Q_1^2}+ \gamma_2\bar Q_1,
\ee
where $\bar Q_1 = v_2, \, \bar P_1 = \frac{v_{2x}- \gamma_0 v_2}{2 v_2^2}$.
\end{subequations}

\paragraph{Defining the $z$ space, using $B_3^z$:}

The stationary $t_n$ flow (from (\ref{wtn-vtn})) is defined by
\begin{subequations}
\be\label{B3zdH=0}
B_3^z \delta_z H_{n}^z=0 \quad\Rightarrow\quad  \delta_z \left(H_n^z-\mu_0 z_0-\mu_1 z_1-\mu_2 z_2\right)=0,
\ee
which gives
\be\label{Lag:B3z}
{\cal L}_n^z = H_n^z-\mu_0 z_0-\mu_1 z_1-\mu_2 z_2.
\ee
For the $t_1$ flow, we use $H_1^z$, removing an exact derivative, to obtain
\be\label{t1-L1z}
{\cal L}_{1}^z =\frac14(z_1^2+z_{2x})^2+z_0z_2z_{2x}+z_0z_1(z_1z_2-2)+z_0^2z_2^2-\mu_0z_0-\mu_1z_1-\mu_2z_2.
\ee
This is degenerate and leads to
\be
z_0=\frac1{36z_2^3}\left((\mu_0z_2+4)\sigma-12z_2^2z_{2x}+\mu_0z_2(10-\mu_0z_2)-6\mu_1z_2^3+8\right),
\quad z_1=\frac{\sigma-\mu_0z_2+2}{6z_2},
\ee
with $\sigma=\sqrt{12\mu_1z_2^3+(\mu_0^2-12z_{2x})z_2^2+8\mu_0z_2+4}$.  The Lagrangian is then written
\be\label{t1-L1z-1}
{\cal L}_{1}^z =-\frac{\sigma^3+8}{108z_2^4}-\frac{2\mu_0}{9z_2^3}+\frac{12(\mu_0z_2+1)z_{2x}-5\mu_0^2}{36z_2^2}+\frac{\mu_0^3-36\mu_1}{108z_2} -\mu_2z_2-\frac13\mu_0\mu_1,
\ee
leading to canonical variables \; $\tilde Q_1=z_2,\; \tilde P_1 = \frac1{6z_2^2}\left(\sigma+2\mu_0z_2+2\right)$\; and Hamiltonian
\be\label{B3-t1-hz}
h^{(\tilde Q)} =-\tilde Q_1^2\tilde P_1^3+ (\mu_0 \tilde Q_1+1)\tilde P_1^2+\frac14(4\mu_1\tilde Q_1-\mu_0^2)\tilde P_1+\mu_2\tilde Q_1.
\ee
\end{subequations}

\subsection{The Miura Maps in these Coordinates}\label{sec:t1-Miura}

We consider the three steps induced by ${\bf w}\mapsto {\bf u}$, ${\bf v}\mapsto {\bf w}$ and ${\bf z}\mapsto {\bf v}$.  We again extend each space to include the parameters as dynamical variables, which we define as ${\bf q} = (q_1,p_1,\alpha_0,\alpha_1,\alpha_2)$, ${\bf \hat Q} = (\hat Q_1,\hat P_1,\hat \beta_0,\hat \beta_1,\hat \beta_2)$, ${\bf \bar Q} = (\bar Q_1,\bar P_1,\gamma_0,\gamma_1,\gamma_2)$ and ${\bf \tilde Q} = (\tilde Q_1,\tilde P_1,\mu_0,\mu_1,\mu_2)$.

\paragraph{The relation of $\bf q$ to $\bf \hat Q$} is given by
\be\label{t1-q1Q1map}
q_1=\hat Q_1,\;\; p_1=\hat Q_1\hat P_1+\frac{\hat \beta_0}2,\;\; \alpha_0 = h^{(\hat Q)}+4\hat \beta_2^2, \;\; \alpha_1=\frac12 \hat \beta_1,\;\;  \alpha_2=\hat \beta_2,
\ee
leading to \;\; $h^{(q)}=\frac18 \hat \beta_0^2$.

\paragraph{The relation of $\bf \hat Q$ to $\bf \bar Q$} is given by
\be\label{t1-Q1P1bmap}
\hat Q_1= \bar Q_1,\;\; \hat P_1= 2 \bar Q_1 \bar P_1\!+\!\frac{2\gamma_1}{\bar Q_1}+\gamma_0,\;\; \hat \beta_0 = \!-4\gamma_1, \;\; \hat \beta_1=\!-2h^{(\bar Q)}\!-\frac12\gamma_0^2,\;\; \hat \beta_2=-\gamma_2,
\ee
leading to \;\; $h^{(\hat Q)}= -2 \gamma_0 \gamma_1$.

\paragraph{The relation of $\bf \bar Q$ to $\bf \tilde Q$} is given by
\be\label{t1-Q1P1tmap}
\bar Q_1=-2\tilde P_1,\;\; \bar P_1= -\frac12\tilde Q_1\tilde P_1+\frac{\mu_2}{4\tilde P_1^2}+\frac14\mu_0,\;\; \gamma_0 = -\mu_1, \;\;
                  \gamma_1=-\mu_2,\quad \gamma_2=-\frac12h^{(\tilde Q)}+\frac14\mu_0\mu_1 ,
\ee
leading to \;\; $h^{(\bar Q)}=\frac12\mu_0\mu_2$.

\subsection{The Array of Poisson Brackets}\label{sec:t1-PiqQQbt}

We constructed the 4 canonical representations \eqref{B0-t1-hu}, \eqref{B1-t1-hw}, \eqref{B2-t1-hv} and \eqref{B3-t1-hz}, respectively on spaces $\bf q$, $\bf \hat Q$, $\bf\bar Q$ and $\bf\tilde Q$.  The canonical brackets are then extended to include the parameters as Casimirs:
\be\label{t1-P0qP1Q}
{\cal{P}}^{(q)}_0={\cal{P}}^{(\hat Q)}_1={\cal{P}}^{(\bar Q)}_2={\cal{P}}^{(\tilde Q)}_3 =
  \left(\begin{array}{rrrrr}
    0&1&0&0&0  \\ -1&0&0&0&0 \\ 0&0&0&0&0 \\ 0&0&0&0&0 \\ 0&0&0&0&0
  \end{array}\right),
\ee
from which we construct the other 12 Poisson brackets, using the Miura maps of Section \ref{sec:t1-Miura} as Poisson maps.

\paragraph{In the ${\bf q}$ space,} (with $h^{(q)}$ given by (\ref{B0-t1-hu})) we have
\begin{subequations}
\bea
&& {\cal{P}}^{(q)}_1 = \begin{pmatrix}
  0&a_{12}& a_{13} &0 &0\\
  -a_{12}&0& a_{23} &0 &0\\
  -a_{13} &-a_{23} &0 &0 &0\\
  0 &0 &0 &0 &0\\
  0 &0 &0 &0 &0
\end{pmatrix} ,
\quad  {\cal{P}}^{(q)}_2   = \begin{pmatrix}
                 0 & b_{12} & 0 & b_{14} & 0\\
                 -b_{12} & 0 & 0 & b_{24} & 0\\
                   0&0&0&0&0\\
                    -b_{14} &-b_{24} &0&0 &0\\
                    0&0&0&0&0
                   \end{pmatrix}, \nn\\[-2mm]
                   &&  \label{Pq-t1}  \\[-2mm]
&&  {\cal{P}}^{(q)}_3   = \begin{pmatrix}
                 0 & c_{12} & 8\alpha_2 c_{15} &0 & c_{15}\\
                 -c_{12} & 0 & 8\alpha_2 c_{25} &0 &  c_{25}\\
                 -8\alpha_2 c_{15} & -8\alpha_2 c_{25} &0&0&0\\
                 0 &0 &0&0 &0\\
                 - c_{15} & -c_{25} & 0&0&0
                   \end{pmatrix} ,  \nn
\eea
where \, $(a_{13},a_{23},0,0,0)^T = (b_{14},b_{24},0,0,0)^T = (c_{15},c_{25},0,0,0)^T = {\cal{P}}^{(q)}_0 \nabla_{q} h^{(q)},\, a_{12}=-h^{(q)}_{\alpha_0},\, b_{12}=-h^{(q)}_{\alpha_1}$,\, $c_{12}=-\left(\pa_{\alpha_2}+8\alpha_2 \pa_{\alpha_0}\right)h^{(q)}$.

\smallskip
Each of these has 3 Casimirs and the $t_1 =t_h$ flow has a quadri-Hamiltonian representation:
\bea
  &&  {\bf q}_{t_h} = {\cal{P}}_0^{(q)} \nabla_q h^{(q)} = {\cal{P}}_1^{(q)} \nabla_q \alpha_0= {\cal{P}}_2^{(q)} \nabla_q \alpha_1= {\cal{P}}_3^{(q)} \nabla_q \alpha_2,   \nn\\
 & &   {\cal{P}}_0^{(q)} \nabla_q \alpha_0 = {\cal{P}}_0^{(q)} \nabla_q \alpha_1= {\cal{P}}_0^{(q)} \nabla_q \alpha_2
  ={\cal{P}}_1^{(q)} \nabla_q h^{(q)} = {\cal{P}}_1^{(q)} \nabla_q \alpha_1={\cal{P}}_1^{(q)} \nabla_q \alpha_2    \label{4Ham-qt1}\\
   &&  ={\cal{P}}_2^{(q)} \nabla_q h^{(q)}={\cal{P}}_2^{(q)} \nabla_q \alpha_0={\cal{P}}_2^{(q)} \nabla_q \alpha_2
   ={\cal{P}}_3^{(q)} \nabla_q h^{(q)}={\cal{P}}_3^{(q)} \nabla_q \alpha_1={\cal{P}}_3^{(q)} \nabla_q(\alpha_0-4\alpha_2^2)=0.\nn
\eea
\end{subequations}

\br
The first 3 rows and columns of ${\cal{P}}^{(q)}_1$ are in the form of a standard $3\times 3$ Poisson bracket, with Casimir $h^{(q)}(q_1,p_1,\alpha_0)$.  The form of ${\cal{P}}^{(q)}_3$ shows how this is generalised, as will be seen in further examples below.
\er

\paragraph{In the ${\bf \hat Q}$ space,} (with $h^{(\hat Q)}$ given by (\ref{B1-t1-hw})) we have
\begin{subequations}
\bea
&& {\cal{P}}^{(\hat Q)}_0 = \frac1{\hat \beta_0}\left(\begin{array}{ccccc}
                         0& a_{12} &  a_{13} &0 &0\\
                         -a_{12} & 0 & a_{23} &0 &0\\
                         - a_{13} & -a_{23} &0&0&0\\
                         0&0&0&0&0\\
                         0&0&0&0&0
                         \end{array}\right),
\quad  {\cal{P}}^{(\hat Q)}_2 =  \begin{pmatrix}
                 0 & b_{12} & 0 & b_{14} &0\\
                  -b_{12} & 0 & 0 & b_{24} &0\\
                   0 & 0 & 0 & 0 &0\\
                     -b_{14} & -b_{24} & 0 & 0 &0\\
                     0 & 0 & 0 & 0 &0
                   \end{pmatrix}, \nn\\[-2mm]
                   &&  \label{PQ-t1}  \\[-2mm]
&&  {\cal{P}}^{(\hat Q)}_3 =  \begin{pmatrix}
                 0 & c_{12} & 0 &0 & c_{15}\\
                  -c_{12} & 0 & 0 &0 & c_{25}\\
                   0 & 0 & 0 & 0 &0\\
                   0 & 0 & 0 & 0 &0\\
                   -c_{15} & -c_{25} & 0 & 0 &0
                   \end{pmatrix} ,  \nn
\eea
where \,$(a_{13},a_{23},0,0,0)^T = (b_{14},b_{24},0,0,0)^T = (c_{15},c_{25},0,0,0)^T ={\cal{P}}^{(\hat Q)}_1 \nabla_{\hat Q} h^{(\hat Q)},\, a_{12}=-h^{(\hat Q)}_{\hat \beta_0},\, b_{12}=-h^{(\hat Q)}_{\hat \beta_1}$, $c_{12}=-h^{(\hat Q)}_{\hat \beta_2}$.

\smallskip
Each of these has 3 Casimirs and the $t_1 =t_h$ flow has a quadri-Hamiltonian representation:
\bea
&&  {\bf \hat Q}_{t_h}= {\cal{P}}_1^{(\hat Q)} \nabla_{\hat Q} h^{(\hat Q)}={\cal{P}}_0^{(\hat Q)} \nabla_{\hat Q} \left(\frac12 \hat \beta_0^2\right) = {\cal{P}}_2^{(\hat Q)} \nabla_{\hat Q} \hat \beta_1
            = {\cal{P}}_3^{(\hat Q)} \nabla_{\hat Q} \hat \beta_2,  \nn \\
&&   {\cal{P}}_1^{(\hat Q)} \nabla_{\hat Q} \hat \beta_0 = {\cal{P}}_1^{(\hat Q)} \nabla_{\hat Q} \hat \beta_1= {\cal{P}}_1^{(\hat Q)} \nabla_{\hat Q} \hat \beta_2
                        ={\cal{P}}_0^{(\hat Q)} \nabla_{\hat Q} h^{(\hat Q)} = {\cal{P}}_0^{(\hat Q)} \nabla_{\hat Q} \hat \beta_1= {\cal{P}}_0^{(\hat Q)} \nabla_{\hat Q} \hat \beta_2    \label{4Ham-Qt1}  \\
&&  ={\cal{P}}_2^{(\hat Q)} \nabla_{\hat Q} h^{(\hat Q)} = {\cal{P}}_2^{(\hat Q)} \nabla_{\hat Q} \hat \beta_0 ={\cal{P}}_2^{(\hat Q)} \nabla_{\hat Q} \hat \beta_2
                                 ={\cal{P}}_3^{(\hat Q)} \nabla_{\hat Q} h^{(\hat Q)} = {\cal{P}}_3^{(\hat Q)} \nabla_{\hat Q} \hat \beta_0= {\cal{P}}_3^{(\hat Q)} \nabla_{\hat Q} \hat \beta_1 = 0.   \nn
\eea
\end{subequations}

\paragraph{In the ${\bf \bar Q}$ space,} (with $h^{(\bar Q)}$ given by (\ref{B2-t1-hv})) we have
\begin{subequations}
\bea
&& {\cal{P}}^{(\bar Q)}_0 = \frac{1}{\gamma_1^2}\left(\begin{array}{ccccc}
                           0& a_{12} & -\gamma_0 a_{13} & \gamma_1 a_{13} &0\\
                            -a_{12} & 0 &  -\gamma_0 a_{23} & \gamma_1 a_{23} &0 \\
                             \gamma_0 a_{13} &  \gamma_0 a_{23} & 0 & 0 &0\\
                              -\gamma_1 a_{13} & -\gamma_1 a_{23} & 0 & 0 &0\\
                              0&0&0&0&0
                             \end{array}\right),
\quad   {\cal{P}}^{(\bar Q)}_1  =  \frac{1}{\gamma_1}\begin{pmatrix}
                           0 & b_{12} & b_{13} & 0 &0\\
                    -b_{12} & 0 & b_{23} & 0 &0\\
                    -b_{13} & -b_{23} & 0 & 0 &0\\
                             0 & 0 & 0 & 0 &0\\
                             0 & 0 & 0 & 0 &0
                             \end{pmatrix}, \nn\\[-2mm]
                   &&  \label{PQb-t1}  \\[-2mm]
&& {\cal{P}}^{(\bar Q)}_3  = \begin{pmatrix}
                 0 & c_{12} & 0 &0 & c_{15}\\
                  -c_{12} & 0 & 0 &0 & c_{25}\\
                   0 & 0 & 0 & 0 &0\\
                   0 & 0 & 0 & 0 &0\\
                   -c_{15} & -c_{25} & 0 & 0 &0
                   \end{pmatrix} ,  \nn
\eea
where \, $(a_{13},a_{23},0,0,0)^T = (b_{13},b_{23},0,0,0)^T = (c_{15},c_{25},0,0,0)^T ={\cal P}_2^{(\bar Q)} \nabla_{\bar Q} h^{(\bar Q)}$ and

\noindent $a_{12}= \left(\gamma_0\pa_{\gamma_0}-\gamma_1\pa_{\gamma_1}\right)\left(h^{(\bar Q)}+\frac{1}{4}\gamma_0^2\right),\, b_{12}=-\pa_{\gamma_0} \left(h^{(\bar Q)}+\frac{1}{4}\gamma_0^2\right),\,
c_{12}=-h^{(\bar Q)}_{\gamma_2}$.

\smallskip
Each of these has 3 Casimirs and the $t_1 =t_h$ flow has a quadri-Hamiltonian representation:
\bea
&&  {\bf \bar Q}_{t_h} = {\cal{P}}_2^{(\bar Q)} \nabla_{\bar Q}h^{(\bar Q)}={\cal{P}}_0^{(\bar Q)} \nabla_{\bar Q} \left(\frac12\gamma_1^2\right) = {\cal{P}}_1^{(\bar Q)} \nabla_{\bar Q}(\gamma_0\gamma_1)
             = {\cal{P}}_3^{(\bar Q)} \nabla_{\bar Q} \gamma_2  ,     \nn   \\
&&  {\cal{P}}_2^{(\bar Q)} \nabla_{\bar Q} \gamma_0 = {\cal{P}}_2^{(\bar Q)} \nabla_{\bar Q} \gamma_1={\cal{P}}_2^{(\bar Q)} \nabla_{\bar Q} \gamma_2
= {\cal{P}}_0^{(\bar Q)}\nabla_{\bar Q}(\gamma_0\gamma_1)= {\cal{P}}_0^{(\bar Q)} \nabla_{\bar Q}\gamma_2= {\cal{P}}_0^{(\bar Q)} \nabla_{\bar Q}\left(h^{(\bar Q)}+\frac{1}{4}\gamma_0^2\right)  \label{4Ham-bQt1}  \\
&& = {\cal{P}}_1^{(\bar Q)}\nabla_{\bar Q}\gamma_1= {\cal{P}}_1^{(\bar Q)} \nabla_{\bar Q}\gamma_2= {\cal{P}}_1^{(\bar Q)} \nabla_{\bar Q}\left(h^{(\bar Q)}+\frac{1}{4}\gamma_0^2\right)
 = {\cal{P}}_3^{(\bar Q)}\nabla_{\bar Q}\gamma_0= {\cal{P}}_3^{(\bar Q)} \nabla_{\bar Q}\gamma_1= {\cal{P}}_3^{(\bar Q)} \nabla_{\bar Q}h^{(\bar Q)}=0.    \nn
\eea
\end{subequations}

\paragraph{In the ${\bf \tilde Q}$ space,} (with $h^{(\tilde Q)}$ given by (\ref{B3-t1-hz})) we have
\begin{subequations}
\bea
&&  {\cal{P}}^{(\tilde Q)}_0 = \frac{1}{\mu_2^3}\left(\begin{array}{ccccc}
                           0& a_{12} & (\mu_1^2-\mu_0\mu_2)a_{13} & -\mu_1\mu_2a_{13} &\mu_2^2a_{13}\\
                            -a_{12} & 0 & (\mu_1^2-\mu_0\mu_2)a_{23} & -\mu_1\mu_2a_{23} &\mu_2^2a_{23}\\
                             (\mu_0\mu_2-\mu_1^2)a_{13} & (\mu_0\mu_2-\mu_1^2)a_{23}  & 0 & 0 &0\\
                              \mu_1\mu_2a_{13} & \mu_1\mu_2a_{23} & 0 & 0 &0\\
                              -\mu_2^2 a_{13}& -\mu_2^2 a_{23} & 0 & 0 &0
                              \end{array}\right),  \nn\\[-2mm]
                              &&  \label{PQt-t1}  \\[-2mm]
&&   {\cal{P}}^{(\tilde Q)}_1  =  \frac{1}{\mu_2^2}\begin{pmatrix}
                           0 & b_{12} & -\mu_1 b_{13} & \mu_2 b_{13} &0\\
                           -b_{12} & 0 & -\mu_1 b_{23} & \mu_2 b_{23} &0\\
                           \mu_1 b_{13} & \mu_1 b_{23} & 0 & 0 &0\\
                            -\mu_2 b_{13} & -\mu_2 b_{23} & 0 & 0 &0\\
                             0 & 0 & 0 & 0 &0
                             \end{pmatrix},   \quad
  {\cal{P}}^{(\tilde Q)}_2  =  \frac{1}{\mu_2}\begin{pmatrix}
                           0 & c_{12} &c_{13} & 0 &0\\
                           -c_{12} & 0 &c_{23} & 0 &0\\
                           -c_{13} & -c_{23} & 0 & 0 &0\\
                            0 & 0 & 0 & 0 &0\\
                            0 & 0 & 0 & 0 &0
                             \end{pmatrix} ,  \nn
\eea
where \, $(a_{13},a_{23},0,0,0)^T = (b_{13},b_{23},0,0,0)^T = (c_{13},c_{23},0,0,0)^T = {\cal P}_3^{(\tilde Q)} \nabla_{\tilde Q} h^{(\tilde Q)}$ and

\noindent $a_{12}= \left((\mu_0\mu_2-\mu_1^2)\pa_{\mu_0} +\mu_1\mu_2 \pa_{\mu_1}-\mu_2^2 \pa_{\mu_2}\right)\left(h^{(\tilde Q)}-\frac{1}{2}\mu_0\mu_1\right),\, b_{12}=\left(\mu_1\pa_{\mu_0}-\mu_2\pa_{\mu_1}\right) \left(h^{(\tilde Q)}-\frac{1}{2}\mu_0\mu_1\right)$, $c_{12}= -\pa_{\mu_0} \left(h^{(\tilde Q)}-\frac{1}{2}\mu_0\mu_1\right)$.

 \smallskip
Each of these has 3 Casimirs and the $t_1 =t_h$ flow has a quadri-Hamiltonian representation:
\bea
&&  {\bf \tilde Q}_{t_h} = {\cal{P}}_3^{(\tilde Q)} \nabla_{\tilde Q}h^{(\tilde Q)}={\cal{P}}_0^{(\tilde Q)} \nabla_{\tilde Q} \left(\frac12\mu_2^2\right) = {\cal{P}}_1^{(\tilde Q)} \nabla_{\tilde Q}(\mu_1\mu_2)
                     ={\cal{P}}_2^{(\tilde Q)} \nabla_{\tilde Q}(\mu_0\mu_2)  ,  \nn\\
&&  {\cal{P}}_3^{(\tilde Q)}\nabla_{\tilde Q}\mu_0= {\cal{P}}_3^{(\tilde Q)} \nabla_{\tilde Q}\mu_1= {\cal{P}}_3^{(\tilde Q)} \nabla_{\tilde Q}\mu_2
                = {\cal{P}}_0^{(\tilde Q)}\nabla_{\tilde Q}(\mu_1\mu_2)= {\cal{P}}_0^{(\tilde Q)} \nabla_{\tilde Q}\left(2\mu_0\mu_2+\mu_1^2\right)   \nn   \\[-1mm]
                   &&    \label{4Ham-tQt1}   \\[-1mm]
&&  = {\cal{P}}_0^{(\tilde Q)} \nabla_{\tilde Q}\left(h^{(\tilde Q)}-\frac{1}{2}\mu_0\mu_1\right)= {\cal{P}}_1^{(\tilde Q)}\nabla_{\tilde Q}\mu_2= {\cal{P}}_1^{(\tilde Q)} \nabla_{\tilde Q}\left(2\mu_0\mu_2+\mu_1^2\right)
            = {\cal{P}}_1^{(\tilde Q)} \nabla_{\tilde Q}\left(h^{(\tilde Q)}-\frac{1}{2}\mu_0\mu_1\right)    \nn\\
&&  \qquad ={\cal{P}}_2^{(\tilde Q)}\nabla_{\tilde Q}\mu_1 = {\cal{P}}_2^{(\tilde Q)}\nabla_{\tilde Q}\mu_2 ={\cal{P}}_2^{(\tilde Q)} \nabla_{\tilde Q}\left(h^{(\tilde Q)}-\frac{1}{2}\mu_0\mu_1\right)=0.  \nn
\eea
\end{subequations}

\br
When $M=3$, we have all 4 spaces, each 5 dimensional and with 4 Poisson brackets.

When $M=2$ (the DWW case), we no longer have the $\tilde {\bf Q}$ space and for each of the other spaces we reduce to 4 dimensions (losing parameters $\alpha_2, \hat\beta_2, \gamma_2$).  For each space we only have the first 3 Poisson brackets, which can be written as $4\times 4$ matrices.

When $M=1$ (the KdV case), we no longer have the $\bar {\bf Q}$ space, with the $\hat {\bf Q}$ and $\bf q$ spaces reducing to 3 dimensions (losing parameters $\alpha_1, \hat\beta_1$).  Each of these spaces only has the first 2 Poisson brackets, which can be written as $3\times 3$ matrices.
\er

\br[Quadri-Hamiltonian formulation of (\ref{B1-t1-N=1})]
Since we have 4 Poisson brackets in the $\bf q$ space, we can use the Poisson map (\ref{PBmap-t1}) to derive 2 further Poisson brackets for (\ref{B1-t1-N=1}), in the $\bf Q$ space.  We don't explicitly present this for the $t_1$ flow, but the corresponding results for the $t_2$ flow are given in Section \ref{sec:quadri-t2}.
\er


\section{The Stationary $t_2$ Flow in Quadri-Hamiltonian Form}\label{sec:quadHt2}

Following the approach outlined in Section \ref{sec:quadHt1}, we now consider the $t_2$ flow, with the same definition of ${\cal{P}}_k^{(m)}$.

\subsection{Defining the Coordinates}

The 4 Poisson matrices ${\cal P}_k^{(k)}$ can be {\em directly constructed} in canonical form, giving us coordinates $(q_i,p_i)$, $(\hat Q_i,\hat P_i)$, $(\bar Q_i,\bar P_i)$ and $(\tilde Q_i,\tilde P_i)$ respectively.

\paragraph{Defining the $u$ space, using $B_0^u$:}

We previously defined these canonical coordinates when deriving (\ref{B0-t2-hq}), giving
\be\label{B0-t2-hu}
h^{(q)} =5q_2p_1^2-4p_1p_2+\alpha_0q_2+\alpha_1(4q_1+q_2^2)-4\alpha_2 q_2 (2q_1+q_2^2)-\frac14q_1^3+\frac3{16}q_1^2q_2^2  +\frac{25}{64}q_1q_2^4+\frac{29}{256}q_2^6.
\ee

\paragraph{Defining the $w$ space, using $B_1^w$:}

Using (\ref{Lag:B1w}), with $n=2$, and choosing $-3H_4^w$, gives
\begin{subequations}
\be\label{t2-L4w}
{\cal L}_{4}^w=\frac38 (4w_1+3w_2^2)(w_0^2+w_{0x})+\frac3{16}w_{2x}^2-\frac3{128}w_2(12w_1+7w_2^2)(4w_1+w_2^2)-\hat \beta_0w_0-\hat \beta_1w_2-\hat \beta_2(4w_1+w_2^2).
\ee
This is degenerate and leads to $w_1=-\frac5{12}w_2^2+\frac2{9w_2}(3w_0^2+3w_{0x}-8\hat \beta_2)$, giving    {\small
\be\label{t2-L4-1}
{\cal L}_{4}^w = \frac{(w_0^2+w_{0x})^2}{2 w_2}+\frac{3w_2^3-16\hat \beta_2}{6w_2}w_{0x}+\frac3{16}w_{2x}^2
                           +\frac1{32}w_2^2(w_2^3+16w_0^2)-\hat \beta_0w_0-\hat \beta_1w_2+\frac29\hat \beta_2\left(3w_2^2-\frac{12w_0^2-16\hat \beta_2}{w_2}\right),
\ee
}corresponding to the Hamiltonian
\be\label{B1-t2-hw}
h^{(\hat Q)} = \frac{1}{2}\hat Q_2\hat P_1^2-\frac16\left(6\hat Q_1^2+3\hat Q_2^3-16\hat \beta_2\right) \hat P_1+\frac43 \hat P_2^2+\frac3{32}\hat Q_2^5+\hat \beta_0 \hat Q_1+\hat \beta_1\hat Q_2-2\hat \beta_2\hat Q_2^2,
\ee
with \, $\hat Q_1=w_0,\; \hat Q_2=w_2,\; \hat P_1=\frac12 w_2^2+\frac1{3w_2}(3w_0^2+3w_{0x}-8\hat \beta_2),\; \hat P_2=\frac38 w_{2x}$.
\end{subequations}

\paragraph{Defining the $v$ space, using $B_2^v$:}

Using (\ref{Lag:B2v}), with $n=2$, and choosing $2H_3^v$ (removing an exact derivative), we obtain
\begin{subequations}
\be\label{t2-L3v}
{\cal L}_{3}^v =\frac12(2v_0v_1+v_{1x})^2-\frac34v_2^2(2v_0v_1+v_{1x})+v_0v_{2x}-v_0^2v_2+\frac5{32}v_2^4-\gamma_0 v_0-\gamma_1 v_1-\gamma_2 v_2.
\ee
This is degenerate and leads to $v_0=\frac{3v_1v_2^2-4v_1v_{1x}-2v_{2x}+2\gamma_0}{4(2v_1^2-v_2)}$.  Removing an exact derivative, we find
\bea
{\cal L}_{3}^v  &=&  \frac1{32(2v_1^2-v_2)}\left(-16v_{1x}^2v_2+8(3v_2^3+4\gamma_0v_1-4v_1v_{2x})v_{1x}-8v_{2x}^2+8(3v_1v_2^2+2\gamma_0)v_{2x}\right.  \nn\\[-1mm]
&& \hspace{2cm} \left. -v_2^4(8v_1^2+5v_2)-8\gamma_0(3v_1v_2^2+\gamma_0)-32(\gamma_1v_1+\gamma_2v_2)(2v_1^2-v_2)\right).    \label{t2-L3v-1}
\eea
The Legendre transformation gives coordinates
\bea
&& \bar Q_1=v_1,\quad \bar Q_2=v_2,\quad \bar P_1=\frac1{4(2v_1^2-v_2)}\left( v_2(3v_2^2-4v_{1x})+4v_1(\gamma_0-v_{2x}) \right),   \nn\\[-2mm]
&&         \label{t2-L3v-QP}             \\[-2mm]
&& \qquad\qquad\quad \bar P_2=\frac1{4(2v_1^2-v_2)}\left( v_1(3v_2^2-4v_{1x})-2v_{2x}+2\gamma_0) \right),  \nn
\eea
and Hamiltonian
\be\label{B2-t2-hv}
h^{(\bar Q)} = \frac12 \bar P_1^2-2\bar Q_1\bar P_1\bar P_2+\bar Q_2\bar P_2^2+\frac34 \bar Q_2^2 \bar P_1+\gamma_0 \bar P_2+\frac18 \bar Q_2^4+\gamma_1 \bar Q_1+\gamma_2 \bar Q_2.
\ee
\end{subequations}

\paragraph{Defining the $z$ space, using $B_3^z$:}

Using (\ref{Lag:B3z}), with $n=2$, and choosing $H_2^z$ (removing an exact derivative), we obtain
\begin{subequations}
\bea
{\cal L}_{2}^z &=&-\frac18(2z_0z_2+z_{2x})^3-\frac38z_1^2z_{2x}^2+\frac12z_{1x}z_{2x}-\frac38 z_1^2(4z_0z_2+z_1^2)z_{2x}+z_0(z_2z_{1x}+z_1z_{2x})-z_0^2    \nn\\
&&   \hspace{2cm}   -\frac18z_1^6-\frac14 z_0z_1(3z_1z_2-4)(2z_0z_2+z_1^2)-\mu_0z_0-\mu_1z_1-\mu_2z_2.     \label{t2-L2z}
\eea
This is degenerate and leads to \, $z_0=\frac1{6z_2^3}\left(2\sigma-2-3z_2^2(z_1^2+z_{2x})+4z_1z_2\right)$, with
$$
\sigma=\sqrt{3z_2^4z_{1x}-3(z_1^3+z_1z_{2x}+\mu_0)z_2^3+(7z_1^2+3z_{2x})z_2^2-4z_1z_2+1},\qquad\mbox{giving}
$$

{\small  \be
{\cal L}_{2}^z = \frac{2(\sigma^3-1)}{27z_2^6}+\frac{4z_1}{9z_2^5}-\frac{11z_1^2}{9z_2^4} +\frac{43z_1^3+9z_1z_{2x}+9\mu_0}{27z_2^3}
                                      -\frac{11z_1^4+10z_1^2z_{2x}+3z_{2x}^2+8\mu_0 z_1}{12z_2^2}+\frac{\mu_0}2\frac{z_1^2}{z_2} -\mu_1z_1-\mu_2z_2.    \label{t2-L2z-1}
\ee
}The Legendre transformation gives coordinates
\be\label{t2-LT}
\tilde Q_1=z_1,\quad \tilde Q_2=z_2,\quad \tilde P_1=\frac{\sigma}{3z_2^2},\quad \tilde P_2=\frac{1-z_1z_2}{3z_2^4}\sigma+\frac{z_1}{3z_2^3}-\frac{5z_1^2+3z_{2x}}{6z_2^2},
\ee
and the Hamiltonian
{\small \bea
\hspace{-1cm}   h^{(\tilde Q)} &=& \tilde P_1^3-\frac{1}{\tilde Q_2^2}\left((\tilde Q_1\tilde Q_2-1) \tilde P_1+ \tilde Q_2^2 \tilde P_2\right)^2+\frac{(3\mu_0-2\tilde Q_1^3)\tilde Q_2^3
      +2\tilde Q_1\tilde Q_2-1}{3\tilde Q_2^4}\tilde P_1+\frac{\tilde Q_1(2-5\tilde Q_1\tilde Q_2)}{3\tilde Q_2}\tilde P_2    \nn \\[-1mm]
\hspace{-1cm}      && +\frac{2}{27\tilde Q_2^6} \left(3 \tilde Q_1^4\tilde Q_2^4 -14 \tilde Q_1^3\tilde Q_2^3 +15 \tilde Q_1^2\tilde Q_2^2 -6 \tilde Q_1\tilde Q_2 +1\right)
      +\frac{\mu_0}6 \frac{4\tilde Q_1\tilde Q_2-3\tilde Q_1^2\tilde Q_2^2-2}{\tilde Q_2^3}+\mu_1\tilde Q_1+\mu_2\tilde Q_2.   \label{B3-t2-hz}
\eea
}
\end{subequations}

\subsection{The Miura Maps in these Coordinates}\label{sec:t2-Miura}

We consider the three steps induced by ${\bf w}\mapsto {\bf u}$, ${\bf v}\mapsto {\bf w}$ and ${\bf z}\mapsto {\bf v}$.  We again extend each space to include the parameters as dynamical variables, which we define as
${\bf q} = (q_i,p_i,\alpha_0,\alpha_1,\alpha_2)$, ${\bf \hat Q} = (\hat Q_i,\hat P_i,\hat \beta_0,\hat \beta_1,\hat \beta_2)$, ${\bf \bar Q} = (\bar Q_i,\bar P_i,\gamma_0,\gamma_1,\gamma_2)$ and ${\bf \tilde Q} = (\tilde Q_i,\tilde P_i,\mu_0,\mu_1,\mu_2)$.

\paragraph{The relation of $\bf q$ to $\bf \hat Q$} is given by
\begin{subequations}
\be\label{t2-q1Q1map}
\begin{split}
&q_1=-\frac34\hat Q_2^2+\frac23\hat P_1,\quad q_2=\hat Q_2,\quad p_1=-\frac23\hat P_2,\quad p_2=-\frac13(\hat Q_1\hat P_1+2\hat Q_2\hat P_2)+\frac{\hat \beta_0}6,\\
&\quad\quad \alpha_0 =-\frac13 h^{(\hat Q)}, \quad \alpha_1=-\frac16\hat \beta_1,\quad   \alpha_2=-\frac13\hat \beta_2,
\end{split}
\ee
leading to \;  $h^{(q)}=-\frac19 f^{(\hat Q)},\;\; f^{(q)}=-\frac1{18}\hat \beta_0^2$, where $f^{(q)}$ is given by \eqref{t2-fq} and
\be\label{B1-t2-fQ-1}
f^{(\hat Q)}=\frac23 \hat P_1^3-\frac32 \hat Q_2^2\hat P_1^2+8\hat Q_1\hat P_1\hat P_2+\left( \frac38 \hat Q_2^4-3\hat Q_1^2\hat Q_2
                   -8\hat \beta_2\hat Q_2+4\hat \beta_1 \right)\hat P_1 -\hat \beta_0 (4 \hat P_2-3\hat Q_1\hat Q_2).
\ee
\end{subequations}

\paragraph{The relation of $\bf \hat Q$ to $\bf \bar Q$} is given by
\begin{subequations}
\bea
&& \hat Q_1=\bar P_2,\quad \hat Q_2=\bar Q_2,\quad \hat P_1= -\frac32\bar P_1,\quad \hat P_2= \frac34(\bar Q_2\bar P_2-\bar Q_1\bar P_1)+\frac38\gamma_0,   \nn\\[-2mm]
&&   \label{t2-Q1P1bmap}  \\[-2mm]
&& \hat  \beta_0 =-\frac32\gamma_1, \quad \hat \beta_1=-\frac34h^{(\bar Q)},\quad \hat \beta_2=-\frac38\gamma_2,  \nn
\eea
leading to \; $h^{(\hat Q)}=\frac32 f^{(\bar Q)}+\frac3{16}\gamma_0^2$ and $f^{(\hat Q)}=\frac94 \gamma_0\gamma_1$, where
\be\label{B2-t2-fv}
f^{(\bar Q)} = {\bar P}_1{\bar P}_2^2+\frac12({\bar Q}_1^2+{\bar Q}_2){\bar P}_1^2+\frac18({\bar Q}_2^3-4\gamma_0{\bar Q}_1+8\gamma_2){\bar P}_1-\gamma_1{\bar P}_2-\frac{\gamma_1}2{\bar Q}_1{\bar Q}_2.
\ee
\end{subequations}

\paragraph{The relation of $\bf \bar Q$ to $\bf \tilde Q$} is given by
\begin{subequations}
\bea
&& \bar Q_1=\tilde Q_1,\quad \bar Q_2=-2\tilde P_1 +\frac{2}{3\tilde Q_2^2}\, \left(1-2 \tilde Q_1\tilde Q_2\right),\quad
\bar P_1=2\tilde P_2+\frac{\mu_0}{\tilde Q_2}-\frac{2}{3\tilde Q_2^4}\, \left(\tilde Q_1\tilde Q_2-1\right)^2,     \nn\\[-2mm]
&&          \label{t2-Q1P1tmap}           \\[-2mm]
&&  \bar P_2=\tilde Q_1\tilde P_1+\tilde Q_2\tilde P_2+\frac{1}{3\tilde Q_2^3}\, \left(\tilde Q_1^2\tilde Q_2^2+\tilde Q_1\tilde Q_2-1\right),
\quad \gamma_0 = 2\mu_1, \quad \gamma_1=2\mu_2,\quad \gamma_2=h^{(\tilde Q)},   \nn
\eea
leading to \; $h^{(\bar Q)}=f^{(\tilde Q)}$ and $f^{(\bar Q)}=\mu_0\mu_2$, where
\bea
f^{(\tilde Q)} &=& 2{\tilde P}_1^2{\tilde P}_2-\frac2{3{\tilde Q}_2^4}({\tilde Q}_1{\tilde Q}_2-1)^2{\tilde P}_1^2 +\frac2{9{\tilde Q}_2^6}(3{\tilde Q}_1^2{\tilde Q}_2^2
                                     -4{\tilde Q}_1{\tilde Q}_2+2)\left(({\tilde Q}_1{\tilde Q}_2-1)^2-3{\tilde Q}_2^4{\tilde P}_2\right){\tilde P}_1  \nn \\
&& +\frac2{9{\tilde Q}_2^4}(1-2{\tilde Q}_1{\tilde Q}_2)(9{\tilde Q}_2^4{\tilde P}_2-3{\tilde Q}_1^2{\tilde Q}_2^2+10{\tilde Q}_1{\tilde Q}_2-5){\tilde P}_2
+\frac{4}{27{\tilde Q}_2^8}({\tilde Q}_1{\tilde Q}_2-1)^2(2{\tilde Q}_1{\tilde Q}_2-1)^2  \nn \\
&& +\frac{\mu_0}{9{\tilde Q}_2^5}\left(9{\tilde Q}_2^4{\tilde P}_1^2 -3{\tilde Q}_2^2(3{\tilde Q}_1^2{\tilde Q}_2^2-4{\tilde Q}_1{\tilde Q}_2+2){\tilde P}_1
                                                    +18(1-{\tilde Q}_1{\tilde Q}_2){\tilde Q}_2^4{\tilde P}_2-11{\tilde Q}_1^2{\tilde Q}_2^2+14{\tilde Q}_1{\tilde Q}_2-5\right)  \nn  \\
&&   \hspace{1cm} +\frac{2\mu_1}{3{\tilde Q}_2^3}\left(3{\tilde Q}_2^4{\tilde P}_2-({\tilde Q}_1{\tilde Q}_2-1)^2\right)
          +\frac{2\mu_2}{3{\tilde Q}_2}({\tilde Q}_1{\tilde Q}_2+1-3{\tilde Q}_2^2{\tilde P}_1) +\frac{\mu_0^2}{2{\tilde Q}_2^2}.   \label{B3-t2-fz}
\eea
\end{subequations}

\subsection{The Array of Poisson Brackets}\label{sec:t2-PiqQQbt}

We constructed the 4 canonical representations, with Hamiltonians \eqref{B0-t2-hu}, \eqref{B1-t2-hw}, \eqref{B2-t2-hv} and \eqref{B3-t2-hz}, respectively on spaces $\bf q$, $\bf \hat Q$, $\bf\bar Q$ and $\bf\tilde Q$.  The canonical brackets are then extended to include the parameters as Casimirs:
\be\label{t2-P0qP1Q}
{\cal{P}}^{(q)}_0={\cal{P}}^{(\hat Q)}_1={\cal{P}}^{(\bar Q)}_2={\cal{P}}^{(\tilde Q)}_3 =
  \left(\begin{array}{rrrrrrr}
    0&0&1&0&0&0&0 \\ 0&0&0&1&0&0&0 \\ -1&0&0&0&0&0&0 \\ 0&-1&0&0&0&0&0 \\ 0&0&0&0&0&0&0 \\ 0&0&0&0&0&0&0\\ 0&0&0&0&0&0&0
  \end{array}\right),
\ee
from which we can construct the other 12 Poisson brackets.  For this paper, we only present the operators ${\cal P}_i^{(q)}$ and ${\cal P}_i^{(\hat Q)}$.

\medskip
In the $\bf q$ space, we have ${\cal{P}}^{(q)}_1$, given by (\ref{t2-P10}), and     
\begin{subequations}\label{t2-Piq}
\be\label{t2-P2q}
{\cal{P}}^{(q)}_2 = \begin{pmatrix}
                         0& 0 & 3q_2^2-4q_1 &\frac{1}{2} q_2(4q_1+15q_2^2) &b_{15} &b_{16} &0\\
                         0& 0 &-4 q_2 &-4q_1- 7q_2^2 &b_{25} &b_{26} &0\\
                         4 q_1-3 q_2^2 & 4q_2 &0 & 4q_2p_1 & b_{35} &b_{36} &0\\
                         -\frac{1}{2} q_2(4q_1+15q_2^2) & 4q_1+ 7q_2^2 &-4q_2p_1 &0 &b_{45} &b_{46} &0\\
                         -b_{15}&-b_{25}&-b_{35}&-b_{45}&0&0&0\\
                         -b_{16}&-b_{26}&-b_{36}&-b_{46}&0&0&0\\
                         0&0&0&0&0&0&0
                         \end{pmatrix},
\ee
\be \label{t2-P3q}
{\cal{P}}^{(q)}_3 = \begin{pmatrix}
                         0& 0& q_2(8q_1+3q_2^2) & c_{14} &0 &c_{16} &c_{17}\\
                         0& 0& -4q_1-q_2^2 &\frac12q_2^3-2q_1q_2&0 &c_{26} &c_{27}\\
                         -q_2(8q_1+3q_2^2) & 4q_1+q_2^2 &0 &p_1(4q_1+q_2^2)&0 &c_{36} &c_{37}\\
                         -c_{14} &2q_1q_2-\frac12q_2^3 &-p_1(4q_1+q_2^2) &0&0&c_{46} &c_{47}\\
                         0&0&0&0&0&0&0\\
                         -c_{16}&-c_{26}&-c_{36}&-c_{46}&0&0&0\\
                         -c_{17}&-c_{27}&-c_{37}&-c_{47}&0&0&0
                         \end{pmatrix},
\ee
where \; $c_{14} = \frac{1}{8}(4 q_1+q_2^2)(4 q_1+15 q_2^2)$, \; and
\bea
&&  \frac{1}{4}(b_{15},b_{25},b_{35},b_{45},0,0,0)^T = (c_{16},c_{26},c_{36},c_{46},0,0,0)^T = {\cal{P}}^{(q)}_0 \nabla_q f^{(q)},\nn\\
&&  (b_{16},b_{26},b_{36},b_{46},0,0,0)^T = 2 (c_{17},c_{27},c_{37},c_{47},0,0,0)^T = {\cal{P}}^{(q)}_0 \nabla_q h^{(q)}.  \nn
\eea
Each of the Poisson matrices has 3 Casimirs and the $t_h$ and $t_f$ flows have quadri-Hamiltonian representations, given below
\bea
   {\bf q}_{t_h} &=& {\cal{P}}_0^{(q)} \nabla_q h^{(q)} = {\cal{P}}_1^{(q)} \nabla_q \left(-\frac{1}{4}\alpha_0\right) = {\cal{P}}_2^{(q)} \nabla_q \alpha_1= {\cal{P}}_3^{(q)} \nabla_q (2\alpha_2), \nn\\
   {\bf q}_{t_f} &=& {\cal{P}}_0^{(q)} \nabla_q f^{(q)} = {\cal{P}}_1^{(q)} \nabla_q \left(-\frac{1}{8} h^{(q)}\right)
                            = {\cal{P}}_2^{(q)} \nabla_q \left(\frac{1}{4}\alpha_0\right) ={\cal{P}}_3^{(q)} \nabla_q \alpha_1,  \nn\\[-2mm]
                            &&              \label{4Ham-q-t2}            \\[-2mm]
  &&  {\cal{P}}_0^{(q)} \nabla_q \alpha_0 = {\cal{P}}_0^{(q)} \nabla_q \alpha_1= {\cal{P}}_0^{(q)} \nabla_q \alpha_2
  ={\cal{P}}_1^{(q)} \nabla_q f^{(q)} = {\cal{P}}_1^{(q)} \nabla_q \alpha_1={\cal{P}}_1^{(q)} \nabla_q \alpha_2    \nn\\[1mm]
   &&  ={\cal{P}}_2^{(q)} \nabla_q h^{(q)}={\cal{P}}_2^{(q)} \nabla_q f^{(q)}={\cal{P}}_2^{(q)} \nabla_q \alpha_2
   ={\cal{P}}_3^{(q)} \nabla_q h^{(q)}={\cal{P}}_3^{(q)} \nabla_q f^{(q)}={\cal{P}}_3^{(q)} \nabla_q \alpha_0=0.    \nn
\eea
\end{subequations}

\smallskip
In the $\bf \hat Q$ space, we have
\begin{subequations}\label{t2-PiQhat}
 {\small  \be
{\cal{P}}^{(\hat Q)}_0 = \frac1{\hat \beta_0}\begin{pmatrix}
                         0 &-4\hat Q_1 &4\hat P_2\!-\!3\hat Q_1\hat Q_2 &-\frac32\hat Q_2\hat P_1\!-\!\frac32\hat Q_1^2\!+\!\frac34\hat Q_2^3\!-\!4\hat \beta_2 &a_{15} &0 &0\\
                         4\hat Q_1& 0&0 &0 &a_{25}  &0&0\\
                         3\hat Q_1\hat Q_2-4\hat P_2&0 &0 &\frac32\hat \beta_0 &a_{35}  &0&0\\
                         \frac32 \hat Q_2\hat P_1\!+\!\frac32\hat Q_1^2\!-\!\frac34\hat Q_2^3\!+\!4\hat \beta_2 &0 &-\frac32\hat \beta_0 &0 &a_{45} &0 &0\\
                         -a_{15} &-a_{25} &-a_{35} &-a_{45} &0 &0 &0\\
                         0&0&0&0&0&0&0 \\
                         0&0&0&0&0&0&0
                         \end{pmatrix}, \label{t2-P0Qhat}
\ee
\be
\hspace{-2.5cm}  {\cal{P}}^{(\hat Q)}_2 = \begin{pmatrix}
                         0 &\frac43 &0 &\hat Q_1 &0 &b_{16} &0\\
                         -\frac43 & 0  &0 &-\hat Q_2 &0 &b_{26} &0\\
                         0 &0 &0 &-\hat P_1 &0 &b_{36} &0\\
                         -\hat Q_1 &\hat Q_2 &\hat P_1 &0 &0 &b_{46} &0\\
                         0 &0 &0 &0 &0 &0 &0\\
                         -b_{16}&-b_{26}&-b_{36}&-b_{46}&0&0&0\\
                         0&0&0&0&0&0&0
                         \end{pmatrix}, \label{t2-P2Qhat}
\ee
\be
{\cal{P}}^{(\hat Q)}_3 = \begin{pmatrix}
                         0& -4\hat Q_2 &-4\hat P_1 &-3\hat Q_1\hat Q_2 &0 &c_{16} &c_{17}\\
                         4\hat Q_2& 0  &0 &3\hat Q_2^2-4\hat P_1&0 &c_{26} &c_{27}\\
                         4\hat P_1 &0 &0 &3\hat Q_2\hat P_1&0 &c_{36} &c_{37}\\
                         3\hat Q_1\hat Q_2 &4\hat P_1-3\hat Q_2^2 &-3\hat Q_2\hat P_1 &0&0&c_{46} &c_{47}\\
                         0&0&0&0&0&0&0\\
                         -c_{16}&-c_{26}&-c_{36}&-c_{46}&0&0&0\\
                         -c_{17}&-c_{27}&-c_{37}&-c_{47}&0&0&0
                         \end{pmatrix},    \label{t2-P3Qhat}
\ee
}where
\bea
&&  (a_{15},a_{25},a_{35},a_{45},0,0,0)^T = (c_{16},c_{26},c_{36},c_{46},0,0,0)^T ={\cal{P}}_1^{(\hat Q)} \nabla_{\hat Q} f^{(\hat Q)},   \nn  \\
&&   (b_{16},b_{26},b_{36},b_{46},0,0,0)^T = \frac{2}{3} (c_{17},c_{27},c_{37},c_{47},0,0,0)^T ={\cal{P}}_1^{(\hat Q)} \nabla_{\hat Q} h^{(\hat Q)} .  \nn
\eea

\medskip
Each of the Poisson matrices has 3 Casimirs and the $t_h$ and $t_f$ flows have quadri-Hamiltonian representations
as below
\bea
 &&  \hspace{-1cm} {\bf \hat Q}_{t_h} = {\cal{P}}_1^{(\hat Q)} \nabla_{\hat Q} h^{(\hat Q)}={\cal{P}}_0^{(\hat Q)} \nabla_{\hat Q} \left(\frac16 f^{(\hat Q)}\right)
                    ={\cal{P}}_2^{(\hat Q)} \nabla_{\hat Q} \hat \beta_1={\cal{P}}_3^{(\hat Q)} \nabla_{\hat Q} \left(\frac23\hat \beta_2\right), \nn\\[-2mm]
&&                \label{4Ham-Qhat-t2-thf}                               \\[-2mm]
 && \hspace{-1cm} {\bf \hat Q}_{t_f} = {\cal{P}}_1^{(\hat Q)} \nabla_{\hat Q} f^{(\hat Q)}={\cal{P}}_0^{(\hat Q)} \left(\frac12\hat \beta_0^2\right)
             ={\cal{P}}_2^{(\hat Q)} \nabla_{\hat Q} (3h^{(\hat Q)})={\cal{P}}_3^{(\hat Q)} \nabla_{\hat Q} \hat \beta_1,  \nn\\[1mm]
 &&  \hspace{-1cm} {\cal{P}}_1^{(\hat Q)} \nabla_{\hat Q} \hat \beta_0 = {\cal{P}}_1^{(\hat Q)} \nabla_{\hat Q} \hat \beta_1= {\cal{P}}_1^{(\hat Q)} \nabla_{\hat Q} \hat \beta_2
                        ={\cal{P}}_0^{(\hat Q)} \nabla_{\hat Q} h^{(\hat Q)}      = {\cal{P}}_0^{(\hat Q)} \nabla_{\hat Q} \hat \beta_1= {\cal{P}}_0^{(\hat Q)} \nabla_{\hat Q} \hat \beta_2     \nn\\[2mm]
  && \hspace{-1cm}  ={\cal{P}}_2^{(\hat Q)} \nabla_{\hat Q} f^{(\hat Q)} = {\cal{P}}_2^{(\hat Q)} \nabla_{\hat Q} \hat \beta_0
               ={\cal{P}}_2^{(\hat Q)} \nabla_{\hat Q} \hat \beta_2 ={\cal{P}}_3^{(\hat Q)} \nabla_{\hat Q} h^{(\hat Q)} = {\cal{P}}_3^{(\hat Q)} \nabla_{\hat Q} f^{(\hat Q)}
                             = {\cal{P}}_3^{(\hat Q)} \nabla_{\hat Q} \hat \beta_0 = 0.   \label{4Ham-Qhat-t2}
\eea
\end{subequations}

\subsection{The Quadri-Hamiltonian Form of (\ref{B1-t2-N=2})}\label{sec:quadri-t2}

The Poisson tensors (\ref{t2-Piq}) of (\ref{B0-t2-hu}) (=(\ref{B0-t2-hq})) were derived by using the Miura maps of Section \ref{sec:t2-Miura}.  However, these Poisson tensors are an intrinsic property of (\ref{B0-t2-hq}), independent of our method of construction, so can be understood in the context of Section \ref{sec:stat-t2}.  We can therefore use the Poisson map (\ref{PBmap-t2}) to construct an additional pair of Poisson brackets for (\ref{B1-t2-N=2}):
\begin{subequations}\small{
\bea
  {\cal{P}}^{(Q)}_2  &=& \begin{pmatrix}
                         0 &0 &2{Q}_1 &{Q}_2 &0&a_{16} &0\\
                         0 &0 &{Q}_2 &0 &0 &a_{26} &0\\
                         -2{Q}_1 &-{Q}_2 &0 &{P}_2 &0&a_{36} &0\\
                         -{Q}_2 &0 &-{P}_2 &0 &0 &a_{46}&0\\
                         0&0&0&0&0&0&0 \\
                         -a_{16} &-a_{26} &-a_{36} &-a_{46} &0 &0 &0\\
                         0&0&0&0&0&0&0
                         \end{pmatrix},     \nn\\[-2mm]
                         &&               \label{t2-P23Q}                               \\[-2mm]
  {\cal{P}}^{(Q)}_3 &=&  \begin{pmatrix}
                         0 &0 &4{Q}_1^2+{Q}_2^2 &2{Q}_1{Q}_2 &0 &b_{16} &b_{17}\\
                         0 & 0  &2{Q}_1{Q}_2 &{Q}_2^2 &0 &b_{26} &b_{27}\\
                         -4{Q}_1^2-{Q}_2^2 &-2{Q}_1{Q}_2 &0 &2{Q}_1 P_2  &0 &b_{36} &b_{37}\\
                         -2{Q}_1{Q}_2 &-{Q}_2^2 &-2{Q}_1{P}_2 &0 &0 &b_{46} &b_{47}\\
                         0&0&0&0&0&0&0\\
                         -b_{16}&-b_{26}&-b_{36}&-b_{46}&0&0&0\\
                         -b_{17}&-b_{27}&-b_{37}&-b_{47}&0&0&0
                         \end{pmatrix},             \nn
\eea
}where \; $(a_{16},a_{26},a_{36},a_{46},0,0,0)^T = -(b_{17},b_{27},b_{37},b_{47},0,0,0)^T =2 {\cal{P}}_1^{(Q)} \nabla_{Q} h^{(Q)}$ and
$(b_{16},b_{26},b_{36},b_{46},0,0,0)^T =2{\cal{P}}_1^{(Q)} \nabla_{Q} f^{(Q)}$, \; where $h^{(Q)}$ and $f^{(Q)}$ are given respectively by (\ref{B1-t2-N=2}) and (\ref{t2-fQ}), for $N=2$.

Each of these has 3 Casimirs:
\be\label{t2-Q-P2P3Cas}
{\cal{P}}_2^{(Q)} \nabla_{Q} f^{(Q)} = {\cal{P}}_2^{(Q)} \nabla_{Q} \beta_2 = {\cal{P}}_2^{(Q)} \nabla_{Q} b_1
              = {\cal{P}}_3^{(Q)} \nabla_{Q} f^{(Q)}  = {\cal{P}}_3^{(Q)} \nabla_{Q} \beta_2  = {\cal{P}}_3^{(Q)} \nabla_{Q} h^{(Q)}.
\ee
Taken together with ${\cal{P}}^{(Q)}_0$ and ${\cal{P}}^{(Q)}_1$ of (\ref{t2-P0q-P0Q}), this renders the flows of $h^{(Q)}$ and $f^{(Q)}$ as quadri-Hamiltonian :
\bea
  {\bf Q}_{t_h} &=& {\cal{P}}_1^{(Q)} \nabla_{Q}h^{(Q)}= {\cal{P}}_0^{(Q)} \nabla_{Q}f^{(Q)}= {\cal{P}}_2^{(Q)} \nabla_{Q} \left(\frac12 b_0\right)
                           = {\cal{P}}_3^{(Q)} \nabla_{Q} \left(-\frac12 b_1\right),   \nn\\[-2mm]
                             &&          \label{t2-Q-lad}                  \\[-2mm]
 {\bf Q}_{t_f}  &=& {\cal{P}}_1^{(Q)} \nabla_{Q}f^{(Q)}= {\cal{P}}_0^{(Q)} \nabla_{Q}(-\beta_2)={\cal{P}}_2^{(Q)} \nabla_{Q} h^{(Q)} ={\cal{P}}_3^{(Q)} \nabla_{Q}\left(\frac12 b_0\right).   \nn
\eea
\end{subequations}

\subsubsection{Recursion Operators}\label{sec:t2-recursion}

For the PDE flows, we build a {\em recursion operator} ${\cal R} = B_1 B_0^{-1}$, which includes the {\em formal inverse} $\pa_x^{-1}$ of the differential operator $\pa_x$.  Then $B_{n+1} = {\cal R} B_n$, which can be {\em formally} continued ad infinitum, but only a finite number of the resulting operators are {\em local}, meaning that they depend upon only {\em differential} operators.  For the $M$ component cKdV systems, only the first $M+1$ operators are locally defined.

For the stationary flows the situation is more complicated, since each Poisson tensor ${\cal P}$ has only rank 4, so cannot be inverted.  However, since the Poisson map (\ref{PBmap-t2}) can be restricted to the 4 components $\phi:(Q_i,P_i)\rightarrow (q_i,p_i)$ (and the restriction is invertible), then the corresponding $4\times 4$ submatrix of ${\cal{P}}^{(Q)}_0$, which we call ${\cal Q}_0^{(Q)}$, is invertible.  Defining ${\cal Q}_1^{(Q)}$ to be the corresponding $4\times 4$ submatrix of ${\cal P}_1^{(Q)}$, we can define a ``recursion operator'':
\begin{subequations}
\be\label{RQ}
 {\cal R}^{(Q)} = {\cal Q}_1^{(Q)} \left({\cal Q}_0^{(Q)}\right)^{-1} =
                           \left(
                            \begin{array}{cccc}
                              2 Q_1 & Q_2 & 0 & 0 \\
                             Q_2 & 0 & 0 & 0 \\
                             0 & P_2 & 2 Q_1 & Q_2 \\
                              -P_2 & 0 & Q_2 & 0 \\
                             \end{array}
                              \right).
\ee
Defining ${\cal Q}_{n+1}^{(Q)}={\cal R}^{(Q)} {\cal Q}_{n}^{(Q)}$, we find {\small
\be\label{Q2Q3Q}
 {\cal Q}_2^{(Q)} = \left(
                     \begin{array}{cccc}
                       0 & 0 & 2 Q_1 & Q_2 \\
                       0 & 0 & Q_2 & 0 \\
                       -2 Q_1 & -Q_2 &  0 & P_2 \\
                       -Q_2 & 0 & -P_2 & 0 \\
                     \end{array}
                   \right), \;\;\;
  {\cal Q}_3^{(Q)} = \left(
                         \begin{array}{cccc}
                           0 & 0 & 4 Q_1^2+Q_2^2 & 2 Q_1 Q_2 \\
                           0 & 0 &  2 Q_1 Q_2 & Q_2^2 \\
                           -4 Q_1^2-Q_2^2 & -2 Q_1 Q_2 &0 & 2 Q_1 P_2 \\
                           -2 Q_1 Q_2 & -Q_2^2 &-2 Q_1 P_2 &0 \\
                         \end{array}
                       \right),
\ee
}which should be compared with the formulae of (\ref{t2-P23Q}).
\end{subequations}

\br[In the ${\bf q}-$space]
We can similarly define ${\cal R}^{(q)} = {\cal Q}_1^{(q)} \left({\cal Q}_0^{(q)}\right)^{-1}$, which also generates ${\cal Q}_n^{(q)}$, for higher Poisson brackets.  In the usual way, these recursion operators are intertwined by the Jacobian $J_\phi$ of the map $\phi$, with ${\cal R}^{(q)} J_\phi = J_\phi {\cal R}^{(Q)}$.
\er

\paragraph{Universality of these formulae:}

The Poisson map (\ref{PBmap-t2}) has a universal character.  The first 4 components (our map $\phi$) are fixed, but $\alpha_0$ and $\beta_2$ must be nonzero, in order to guarantee that $h^{(Q)}$ and $f^{(q)}$ are nontrivial.  This minimal requirement is satisfied by the hierarchies for all values of $M$ in (\ref{zero-c}), with the minimal achieved by $M=1$ (the KdV case).  As we increase $M$, then we add more parameters, with the corresponding relations, as seen in the case of $M=2$ and $M=3$.  We \underline{conjecture} that the formula for ${\cal Q}_n^{(Q)}$ represents this part of ${\cal P}_n^{(Q)}$, for all $n$.

The $M$ component cKdV hierarchy has $M+1$ \underline{local} Hamiltonian operators, but the modifications have a diminishing number, as depicted in Figure \ref{B3B2B1B0-fig}.  However, \underline{each} stationary hierarchy has $M+1$ {\em local} Poisson brackets $P_k^{(m)}$, as described at the beginning of Section \ref{sec:quadHt1}.  Unlike the Poisson map (\ref{PBmap-t2}), the corresponding one between $\bf \hat Q$ and $\bf q$ cannot be restricted to the first 4 components, since $\hat \beta_0$ enters the formulae.  This leaves a vestige of nonlocality in the formulae for $P_k^{(m)}$, with $k<m$, in that $\hat \beta_j$ appear explicitly in the $4\times 4$ submatrices.  The corresponding recursion operators are given by ${\cal R} = {\cal Q}_{m+1}^{(m)} \left({\cal Q}_m^{(m)}\right)^{-1}$ and only connect ${\cal Q}_{k}^{(m)}$, for $k\geq m$.


\section{The Poisson Brackets of the General Parabolic Potentials}\label{sec:genPoisson}

We see that the structure of the Poisson bracket $P_0^{(Q)}$ for the $t_2$ flow in the KdV and DWW cases (see \cite{f23-1}) and (\ref{t2-P0q-P0Q}) of this paper have identical form (up to additional zero rows and columns), depending upon the function $f^{(Q)}$. The Poisson bracket $P_2^{(Q)}$ is similarly common to the DWW case and (\ref{t2-P23Q}) of this paper.  In all these cases, the Hamiltonian $h^{(Q)}$ belongs to the standard series known to be separable in parabolic coordinates (see, again, Equation 2.2.41 in \cite{90-16}).

Here we consider the general case of potential $U(Q_1,Q_2)$, separable in parabolic coordinates.  In all our examples (such as (\ref{B1-t2-N=2})) the potential $U(Q_1,Q_2)$ depends upon some additional parameters in a specific way.  These parameters arose as Casimirs of the (degenerate) {\em canonical bracket}, whilst their coefficients are mainly related to the Casimirs of the original Poisson brackets of the corresponding PDE.

For our initial calculation of the general form of $U(Q_1,Q_2)$, we suppress this parametric dependence, but when we consider additional Poisson brackets these parameters explicitly appear as an essential ingredient.

\subsection{Separation Coordinates}

The existence of a pair of commuting integrals, which are {\em quadratic} in momenta, means that we can {\em simultaneously diagonalise} the quadratic parts by constructing separation variables.  Generalising our $h^{(Q)}$ and $f^{(Q)}$, consider
\begin{subequations}
\be\label{genParab-hQfQ}
 h^{(Q)} = \frac{1}{2} (P_1^2+P_2^2)+U(Q_1,Q_2),\quad   f^{(Q)} = P_2 \, (Q_2 P_1-Q_1 P_2) +W(Q_1,Q_2),
\ee
where $U$ and $W$ take specific forms in our examples, but here are considered arbitrary. Using the canonical Poisson bracket $P_1^{(Q)}$, we find
\be\label{genParab-fP1h}
\{f^{(Q)},h^{(Q)}\}_1 = 0 \quad\Rightarrow\quad W_{Q_1} =  Q_2 U_{Q_2},\;\;\; W_{Q_2} = Q_2 U_{Q_1}-2 Q_1 U_{Q_2} ,
\ee
whose integrability condition implies a second order, linear hyperbolic equation for $U$:
\be\label{genParab-Ueq}
Q_2 \left(U_{Q_1Q_1}-U_{Q_2Q_2}\right) - 2 Q_1 U_{Q_1Q_2} = 3 U_{Q_2}.
\ee
The characteristic coordinates for this equation (equivalently, the {\em separation coordinates} for the pair (\ref{genParab-hQfQ})) are
\be\label{genParab-Ueq-uv}
u = \sqrt{Q_1^2+Q_2^2}+Q_1,\;\;  v = \sqrt{Q_1^2+Q_2^2}-Q_1 \;\;\;\Rightarrow\;\;\; \left((u+v) U\right)_{uv}=0,
\ee
leading to
\be\label{genParab-Wuv}
U = \frac{A_1(u)+A_2(v)}{u+v} \quad\mbox{and}\quad W = \frac{v A_1(u)-u A_2(v)}{u+v}.
\ee
\end{subequations}

\subsection{The Jacobi Identity for Non-Canonical Brackets}

The non-canonical brackets have nontrivial entries in the rows and columns corresponding to the Casimirs of the canonical bracket.  The equations of the {\em Jacobi identity} therefore include derivatives with respect to these parameters.  These can be solved to get explicit dependence of $U$ and $W$ on the parameters, which are given the generic names $\kappa_i$ in this section.

\smallskip
For these calculations, it is more convenient to use the general form with $U(Q_1,Q_2,\kappa_1,\kappa_2,\kappa_3)$ and $W(Q_1,Q_2,\kappa_1,\kappa_2,\kappa_3)$, depending upon parameters $\kappa_i$, associated with the last 3 columns of the Poisson brackets $P_k^{(Q)}$.

\paragraph{The Jacobi Identity for $P_0^{(Q)}$:}

Since this Poisson matrix (see (\ref{t2-P0q-P0Q})) only has additional entries in the $\kappa_1$ column, the Jacobi identity just implies constraints on the $\kappa_1$ dependence of both $U$ and $W$:
\begin{subequations}
\be\label{genParab-k1-eq}
W_{Q_1\kappa_1}+\frac{2}{Q_2^2} = W_{Q_2\kappa_1}-\frac{4Q_1}{Q_2^3}=0\quad\Rightarrow\quad \left\{ \begin{array}{l}
                                                                                                    W=-\frac{2\kappa_1Q_1}{Q_2^2}+W_1(Q_1,Q_2,\kappa_2,\kappa_3), \\[1.5mm]
                                                                                                    U = \frac{\kappa_1}{Q_2^2} + U_1(Q_1,Q_2,\kappa_2,\kappa_3),
                                                                                                    \end{array}   \right.
\ee
with the formula for $U$ being derived by using (\ref{genParab-fP1h}).

\paragraph{The Jacobi Identity for $P_2^{(Q)}$:}

Since this Poisson matrix (see (\ref{t2-P23Q})) only has additional entries in the $\kappa_2$ column, the Jacobi identity just implies constraints on the $\kappa_2$ dependence of both $U$ and $W$:
\be\label{genParab-k2-eq}
U_{Q_1\kappa_2}+1 = U_{Q_2\kappa_2}=0                                \quad\Rightarrow\quad \left\{ \begin{array}{l}
                                                                                                    U_1=-\kappa_2 Q_1+U_2(Q_1,Q_2,\kappa_3), \\[1.5mm]
                                                                                                    W_1 = -\frac{1}{2} \kappa_2 Q_2^2 + W_2(Q_1,Q_2,\kappa_3),
                                                                                                    \end{array}   \right.
\ee
with the formula for $W$ being derived by using (\ref{genParab-fP1h}).

\paragraph{The Jacobi Identity for $P_3^{(Q)}$:}

Since this Poisson matrix (see (\ref{t2-P23Q})) has additional entries in {\em both} the $\kappa_2$ and $\kappa_3$ columns, the Jacobi identity implies constraints on {\em both} the $\kappa_2$ and $\kappa_3$ dependence of both $U$ and $W$, but the $\kappa_2$ equations are {\em identically satisfied}. We are left with some equations:
\be\label{genParab-k3-eq}
\left.  \begin{array}{l}
         U_{Q_1\kappa_3}-4 Q_1 = U_{Q_2\kappa_3}-Q_2 = 0 \\[1.5mm]
         W_{Q_1\kappa_3}- Q_2^2 = W_{Q_2\kappa_3}- 2 Q_1 Q_2 =0
         \end{array}     \right\}                         \quad\Rightarrow\quad \left\{ \begin{array}{l}
                                                                                                    U_2= \frac{1}{2}\kappa_3 (4 Q_1^2+Q_2^2)+U_3(Q_1,Q_2), \\[1.5mm]
                                                                                                    W_2 =  \kappa_3 Q_1 Q_2^2 + W_3(Q_1,Q_2),
                                                                                                    \end{array}   \right.
\ee
giving
\bea
 h^{(Q)} &=&  \frac{1}{2} (P_1^2+P_2^2)+\frac{A_1(u)+A_2(v)}{2 \sqrt{Q_1^2+Q_2^2}} +\frac{\kappa_1}{Q_2^2} -\kappa_2 Q_1 + \frac{1}{2} \kappa_3 (4 Q_1^2+Q_2^2) ,\nn\\[-2mm]
 &&   \label{genParab-hQfQ-solk123}  \\[-2mm]
   f^{(Q)} &=&  P_2 \, (Q_2 P_1-Q_1 P_2)+\frac{v A_1(u)-u A_2(v)}{2 \sqrt{Q_1^2+Q_2^2}} -\frac{2 \kappa_1 Q_1}{Q_2^2} -\frac{1}{2} \kappa_2 Q_2^2+\kappa_3 Q_1 Q_2^2,  \nn
\eea
where $u,\, v$ are given by (\ref{genParab-Ueq-uv}).  We see that (\ref{B1-t2-N=2}) and (\ref{t2-fQ}) (with $N=2$) have precisely this form, but with a specific choice of the functions $A_i$ and $(\kappa_1,\kappa_2,\kappa_3)=(\beta_2,b_0,b_1)$.
\end{subequations}

\br
Curiously, the {\em same} functional coefficients of $\kappa_i$ have arisen in two seemingly {\em different} ways: here they were a consequence of the Jacobi identity, whilst previously they arose as Casimirs of the Hamiltonian structures in the PDE context.
\er

\br
These $\kappa_i$ terms correspond to
$$
A_1^{\kappa}(u) =\frac{\kappa_1}{u}-\frac{1}{2} \kappa_2 u^2 + \frac{1}{2} \kappa_3 u^3,\quad A_2^{\kappa}(v)=\frac{\kappa_1}{v} + \frac{1}{2} \kappa_2 v^2 +\frac{1}{2} \kappa_3 v^3.
$$
\er
In summary, we have
\bp
The general system (\ref{genParab-hQfQ-solk123}) is quadri-Hamiltonian, with Poisson brackets $P_k^{(Q)}$, for $k=0, \dots ,3$.  The reduction with $\kappa_3=0$ is tri-Hamiltonian, whilst that with $\kappa_2=\kappa_3=0$ is bi-Hamiltonian.
\ep

\subsection{The Poisson Map Between $\bf q$ and $\bf Q$ Variables}

The Poisson map (\ref{PBmap-t2}) (with a relabelling of parameters), can be used in this more general context. The formulae are dictated purely by the ``kinetic'' parts of the various functions, not the form of the potential.  We then find
\begin{subequations}
\bea
h^{(q)} &=& 5q_2p_1^2-4p_1p_2 +\frac{\bar u A_1(\bar u)-\bar vA_2(\bar v)}{16(\bar u+\bar v)} +\alpha_0q_2+\alpha_1(4q_1+q_2^2)-4\alpha_2 q_2 (2q_1+q_2^2),   \nn\\[-2mm]
&&        \label{Gen-hqfq}              \\[-2mm]
f^{(q)} &=& \frac{1}{4} \left(4 q_1-5q_2^2\right)p_1^2+4q_2p_1p_2-2p_2^2-\frac{\bar u \bar v (A_1(\bar u)+ A_2(\bar v))}{64 (\bar u+\bar v)}  \nn\\[-1mm]
   &&   \hspace{3cm}   +\frac{1}{4}\alpha_0\left(4 q_1+3 q_2^2\right) -\frac{1}{2}\alpha_1q_2\left(4q_1+3q_2^2\right) -\frac{1}{4}\alpha_2(4 q_1+q_2^2)(4 q_1+3 q_2^2),    \nn
\eea
where
\be
\bar u=\sqrt{-8q_1-5q_2^2}+q_2,\quad \bar v=\sqrt{-8q_1-5q_2^2}-q_2, \quad \kappa_2 = 16 \alpha_1, \quad \kappa_3=8 \alpha_2.
\ee
\end{subequations}
Again, the coefficients of $\alpha_i$ are the ones required in order that the Poisson brackets (\ref{t2-Piq}) satisfy the Jacobi identity.  Again, $h^{(q)}$ and $f^{(q)}$ {\em commute} with respect to each of these Poisson brackets.

\br
The specific case of (\ref{B0-t2-hq}) and (\ref{t2-fq}) corresponds to $A_1=\frac{1}{128} \bar u^6,\; A_2= -\frac{1}{128} \bar v^6$.
\er

\subsection{Compatibility with the Lax Representation}\label{sec:compat-Lax}

We now consider the constraints on $h^{(Q)}$ and $f^{(Q)}$ imposed by the existence of the Lax matrix (\ref{L2-Q}), whose $a_{21}$ component is given by (\ref{L2-hQfQ-a21}), but with $h^{(Q)}$ and $f^{(Q)}$ given by (\ref{genParab-hQfQ}).  The term $\frac{\pa f^{(Q)}}{\pa Q_1}$ includes $W_{Q_1}$, but this can be written as $Q_2 U_{Q_2}$, by using (\ref{genParab-fP1h}).
The characteristic equation of the resulting $L^{(2)}$ is of the form
\begin{subequations}
\be\label{L2-gen-hQ-char}
z^2 + \sum_{i=0}^3 \chi_i \lambda^i+ 256 \kappa_3 \lambda^4- 256\lambda^7=0,
\ee
where each coefficient must be a {\em constant} (of motion). In particular,
\be \label{L2-gen-hQ-chi03}
\chi_0 = -Q_2^3 U_{Q_2} + \frac{1}{16} Q_1Q_2^4 (8 Q_1^2+3 Q_2^2) + \kappa_3 Q_2^4, \quad \chi_3 = 80 Q_1^4 + 48 Q_1^2Q_2^2 + 3 Q_2^4 + 256 \kappa_3 Q_1 -64 U_{Q_1},
\ee
so setting $\chi_0=2\kappa_1$ and $\chi_3 = 64 \kappa_2$ gives equations for $U_{Q_i}$, which are compatible and have solution
\be\label{L2-gen-hQ-Usol}
U(Q_1,Q_2) = \frac{\kappa_1}{Q_2^2} - \kappa_2 Q_1 +\frac{1}{2} \kappa_3 (4 Q_1^2+Q_2^2) +\frac{1}{64} Q_1 (4 Q_1^2+Q_2^2)(4 Q_1^2+3 Q_2^2),
\ee
reproducing the form of $h^{(Q)}$ in (\ref{B1-t2-N=2}) (with $(\kappa_1,\kappa_2,\kappa_3)=(\beta_2,b_0,b_1)$).  With these formulae, the characteristic equation (\ref{L2-gen-hQ-char}) reduces to
\be\label{L2-gen-hQ-char1}
z^2 +2\kappa_1 +8 f^{(Q)} \lambda -32 h^{(Q)} \lambda^2+64 \kappa_2 \lambda^3+ 256 \kappa_3 \lambda^4- 256\lambda^7=0,
\ee
which is just (\ref{L2-char-Q}), with $N=2$.
\end{subequations}

\br
We can easily add lower degree terms, corresponding to the stationary KdV and DWW flows.  Clearly we cannot expect to represent the general case of (\ref{genParab-hQfQ-solk123}) in terms of this class of Lax matrix.
\er

\subsection{The Sequence of Polynomial Potentials}\label{sec:Lax-recurs}

Whilst the Poisson brackets allow {\em any} of the potentials in (\ref{genParab-hQfQ-solk123}), we have seen that the Lax matrix (\ref{L2-Q}) fixes the potential to be (\ref{L2-gen-hQ-Usol}).  The {\em homogeneous polynomial} forms of (\ref{genParab-Wuv}), of degree $n$, are given by the choice $A_1(u)=u^{n+1},\, A_2(v) = (-1)^n v^{n+1}$ and it is these particular potentials which appear in stationary reductions of the $M$ component cKdV hierarchy, corresponding to $M=n-2$.

In the Lax matrix (\ref{L2-Q}), the only element which depends upon $M$ is the $a_{21}$ component (\ref{L2-hQfQ-a21}), which can be written
\begin{subequations}
\be\label{a21-M}
a_{21}^{(M)} = -u^{(M)}\, \left(16 \lambda^2+8 Q_1 \lambda-Q_2^2\right) + 4 \frac{\pa h^{(Q)}}{\pa Q_1} \lambda - \frac{\pa f^{(Q)}}{\pa Q_1},\;\;\mbox{where}\;\; u^{(M)} = \sum_{i=0}^{M-1} u_i \lambda^i -\lambda^M,
\ee
with each $u_i$ replaced by an appropriate function of $Q_1, Q_2$.  Going through the stationary reduction route to obtain these formulae for $u_i$ is laborious.  However, it is possible to divorce the construction of $L^{(2)}$ from the stationary reduction process and recursively build $a_{21}^{(M)}$.  In fact, we can recursively build $u^{(M)}$, with the formula
\be\label{uMtoM+1}
u^{(M+1)} = \lambda\, u^{(M)} + g_{M+1}(Q_1,Q_2),
\ee
where $u^{(M)}$ is {\em known}, and a single {\em unknown} function $g_{M+1}(Q_1,Q_2)$ is to be determined, along with the new $h^{(Q)}$ and $f^{(Q)}$. We follow the approach of Section \ref{sec:compat-Lax}, using the characteristic equation of $L^{(2)}$, with $u^{(M+1)}$.  To illustrate the procedure, consider the Lax matrix (\ref{L2-Q}), with
\be\label{M=3to4}
u^{(3)} = \frac{1}{16}Q_1(8Q_1^2+3 Q_2^2)+\kappa_3 -\frac{1}{8}(6 Q_1^2+Q_2^2)\lambda + Q_1 \lambda^2 - \lambda^3,\;\;\;\mbox{and}\;\; u^{(4)} = \lambda\, u^{(3)} + g_{4}(Q_1,Q_2).
\ee
The characteristic equation of the resulting $L^{(2)}$, is
\be\label{L2-M+1-char}
z^2 + \sum_{i=0}^4 \chi_i \lambda^i + 256\kappa_3 \lambda^5- 256\lambda^8=0,
\ee
with
\bea
\chi_4 &=& 80Q_1^4+48Q_1^2Q_2^2+3Q_2^4+256 \kappa_3 Q_1+256g_4,\nn\\
\chi_3 &=& -64U_{Q_1}+Q_1(32Q_1^4+8Q_1^2Q_2^2-3Q_2^4)+32 \kappa_3 (2Q_1^2-Q_2^2)+256Q_1g_4, \label{X430} \\
\chi_0 &=& Q_2^3\left(Q_2g_4-U_{Q_2}\right).  \nn
\eea
Setting $\chi_4= 256 \kappa_4,\, \chi_3 = 64 \kappa_2,\, \chi_0= 2 \kappa_1$, we find
\bea
g_4(Q_1,Q_2) &=& \kappa_4-\kappa_3 Q_1 -\frac{1}{256} \left(80Q_1^4+48Q_1^2Q_2^2+3Q_2^4\right),  \label{g-sol}  \\
U(Q_1,Q_2) &=& \frac{\kappa_1}{Q_2^2} -\kappa_2 Q_1 +\frac{1}{2} \kappa_4 (4 Q_1^2+Q_2^2) -\frac12 \kappa_3Q_1(2Q_1^2+Q_2^2)  \nn\\[-1mm]
  &&   \hspace{4.5cm} -\frac{1}{512} \left(64 Q_1^6+ 80 Q_1^4Q_2^2+24 Q_1^2Q_2^4+Q_2^6\right).
\eea
\end{subequations}

\br
The Lax equations for $L^{(2)}$, with $a_{21}^{(M)}$ are just (\ref{L2x-phi}), with
$$
L^{(1)} = \left(\begin{array}{cc}
              P_1 & -2(2\lambda + Q_1) \\
              b_{21} & -P_1
\end{array}\right),
\quad U = \left(\begin{array}{cc}
0 & 1 \\
-u^{(M)} & 0
\end{array}\right),
$$
where $b_{21} = 2(2\lambda + Q_1) u^{(M)} -\frac{\pa h^{(Q)}}{\pa Q_1}$.
\er


\section{Conclusions}

In this paper we have considered the general structure of the $t_1$ and $t_2$ stationary flows of the $M$ component cKdV hierarchy, giving detailed formulae for the case $M=3$.  One of our reductions, using a squared eigenfunction representation, gives an $N$ component, superintegrable system with a Lax pair.  For $N=2$, we presented Poisson maps which led to multi-Hamiltonian representations.

The $t_2$ stationary flow is separable in parabolic coordinates and the Poisson brackets were generalised to include the entire class of separable potentials.  The Jacobi identity imposed a specific dependence on the parameters $\kappa_i$.  The recursive procedure of Section \ref{sec:Lax-recurs} allows us to build the Lax matrices without any reference to the stationary reduction. Similarly, the ``recursion operator'' of Section \ref{sec:t2-recursion} allows us to build higher Poisson tensors, without needing to go through the entire stationary reduction process.

An important open problem is to generalise these Poisson brackets to the case $N\geq 3$.  In particular, for $N=3$ it would be interesting to find multi-Hamiltonian formulations for the generalisations given in Section 6 of \cite{f23-1}, which are related to generalised parabolic coordinates and also the rational Calogero-Moser model.  This connection first arose in \cite{f22-1}, in which we also presented a conformally flat extension.  Neither a Lax representation nor a multi-Poisson formulation are known for this case.

The matrices of (\ref{zero-c}) have a polynomial dependence on $\lambda$.  In \cite{f87-5,f89-2} we gave a more general formulation, with rational dependence on $\lambda$, which includes the Ito hierarchy.  Furthermore, the latter includes both {\em positive} and {\em negative} hierarchies.  It would be interesting to know whether further cases of the general parabolic potentials of Section \ref{sec:genPoisson} can be given a Lax representation in this way.

Stationary flows are related to the $t$ translation symmetry of the corresponding PDE.  Using the scaling symmetry, it is possible to reduce to Painlev\'e type equations (see \cite{98-4,01-7}).  Such reductions would be interesting to investigate.

\subsection*{Acknowledgements}

This work was supported by the National Natural Science Foundation of China (grant no. 11871396).


\begin{thebibliography}{10}

\bibitem{f87-5}
M.~Antonowicz and A.P. Fordy.
\newblock Coupled {KdV} equations with multi-{Hamiltonian} structures.
\newblock {\em Physica D}, 28:345--57, 1987.

\bibitem{f89-2}
M.~Antonowicz and A.P. Fordy.
\newblock Factorisation of energy dependent {Schr\"odinger} operators: {Miura}
  maps and modified systems.
\newblock {\em Comm. Math. Phys.}, 124:465--86, 1989.

\bibitem{f87-3}
M.~Antonowicz, A.P. Fordy, and S.~Wojciechowski.
\newblock Integrable stationary flows : {Miura} maps and bi-{Hamiltonian}
  structures.
\newblock {\em Phys. Lett. A}, 124:143--50, 1987.

\bibitem{f95-3}
S.~Baker, V.Z. Enolskii, and A.P. Fordy.
\newblock Integrable quartic potentials and coupled {KdV} equations.
\newblock {\em Phys. Lett. A}, 201:167--74, 1995.

\bibitem{09-10}
A.~Ballesteros, A.~Enciso, F.J. Herranz, and O.~Ragnisco.
\newblock Superintegrability on {N-dimensional} curved spaces: {Central}
  potentials, centrifugal terms and monopoles.
\newblock {\em Ann. Phys.}, 324:1219--33, 2009.

\bibitem{76-5}
O.I. Bogoyavlenskii and S.P. Novikov.
\newblock The relationship between {Hamiltonian} formalisms of stationary and
  nonstationary problems.
\newblock {\em Funct. Anal. Appl.}, 10:8--11, 1976.

\bibitem{14-2}
M.~Cariglia.
\newblock Hidden symmetries of dynamics in classical and quantum physics.
\newblock {\em Rev. Modern. Phys.}, 86:1283--1333, 2014.

\bibitem{90-22}
N.W. Evans.
\newblock Superintegrability in classical mechanics.
\newblock {\em Phys. Rev. A}, 41:5666--76, 1990.

\bibitem{00-4}
G.~Falqui, F.~Magri, M.~Pedroni, and J.P. Zubelli.
\newblock A bi-{Hamiltonian} theory for stationary {KdV} flows and their
  separability.
\newblock {\em Regul. Chaotic Dyn.}, 5:33--52, 2000.

\bibitem{f91-1}
A.P. Fordy.
\newblock The {H\'enon-Heiles} system revisited.
\newblock {\em Physica D}, 52:201--210, 1991.

\bibitem{f22-1}
A.P. Fordy and Q.~Huang.
\newblock Integrable and superintegrable extensions of the rational
  {Calogero-Moser} model in three dimensions.
\newblock {\em J. Phys. A}, 55:225203 (36 pages), 2022.

\bibitem{f23-1}
A.P. Fordy and Q.~Huang.
\newblock Stationary flows revisited.
\newblock {\em SIGMA}, 19:015 (34 pages), 2023.

\bibitem{98-4}
A.N.W. Hone.
\newblock Non-autonomous {H\'enon-Heiles} systems.
\newblock {\em Physica D}, 118:1--16, 1998.

\bibitem{01-7}
A.N.W. Hone.
\newblock Coupled {Painlev\'e} systems and quartic potentials.
\newblock {\em J.Phys.A}, 34:2235--45, 2001.

\bibitem{13-2}
W.~Miller Jr, S.~Post, and P.~Winternitz.
\newblock Classical and quantum superintegrability with applications.
\newblock {\em J. Phys. A}, 46:423001, 97 pages, 2013.

\bibitem{90-16}
A.~M. Perelomov.
\newblock {\em Integrable system of classical mechanics and {Lie} algebras}.
\newblock Birkh{\"a}user, Basel, 1990.

\end{thebibliography}

\end{document}